\documentclass[referee, a4paper]{raa}           

\usepackage{graphicx,times}
\usepackage{natbib}
\usepackage{amssymb,amsmath}
\bibpunct{(}{)}{;}{a}{}{,}

\usepackage[pagebackref=true]{hyperref}

\begin{document}

   \title{Spatial distribution of HOCN around Sagittarius B2}

 \volnopage{ {\bf 20XX} Vol.\ {\bf X} No. {\bf XX}, 000--000}
   \setcounter{page}{1}

   \author{Si-Qi Zheng 
   \inst{1,2,3}, Juan Li\inst{1,2}, Jun-Zhi Wang\inst{1,2}, Feng Gao
      \inst{4},  Ya-Jun Wu\inst{1,2},  Shu Liu\inst{5}, Shang-Huo Li\inst{6}
   }

   \institute{ Shanghai Astronomical Observatory, Chinese Academy of Sciences, No. 80 Nandan Road, Shanghai, 200030, China; {\it zhengsq@shao.ac.cn, lijuan@shao.ac.cn, jzwang@shao.ac.cn}\\
        \and
             Key Laboratory of Radio Astronomy, Chinese Academy of Sciences,  10 Yuanhua Road, Nanjing, JiangSu 210033, PR China\\
	\and
%	  Center for Astrophysics, University of Science and Technology of China, Hefei 230026, China\\
University of Chinese Academy of Sciences, No.19(A) Yuquan Road, Shijingshan District, Beijing, P.R.China 100049\\
\and
Hamburger Sternwarte, Universitaet Hamburg, Gojenbergsweg 112, 21029, Hamburg, Germany\\
\and
National Astronomical Observatories, Chinese Academy of Sciences, Beijing 100101, China\\
\and 
Korea Astronomy and Space Science Institute, 776 Daedeokdae-ro, Yuseong-gu, Daejeon 34055, Republic of Korea\\
\vs \no
   {\small Received 20XX Month Day; accepted 20XX Month Day}
}

\abstract{HOCN and HNCO abundance ratio in molecular gas can tell us the information of their formation mechanism. We performed high-sensitivity mapping observations of HOCN, HNCO, and HNC$^{18}$O lines  around Sagittarius B2 (Sgr B2) with IRAM 30m telescope at 3-mm wavelength.   HNCO 4$_{04}$-3$_{03}$ and HOCN 4$_{04}$-3$_{03}$ are used to obtain the abundance ratio of HNCO to HOCN. The ratio of HNCO 4$_{04}$-3$_{03}$ to HNC$^{18}$O 4$_{04}$-3$_{03}$ is used to calculate the optical depth of HNCO 4$_{04}$-3$_{03}$.  The abundance ratio of HOCN and HNCO is observed to range from 0.4\% to 0.7\% toward most positions, which agrees well with the gas-grain model. However, the relative abundance of HOCN is observed to be enhanced toward the direction of Sgr B2 (S), with HOCN to HNCO abundance ratio of $\sim$ 0.9\%. The reason for that still needs further investigation. %From the spatial distribution of their emission, HOCN and HNCO  are  found to be  extended  toward Sgr B2. HOCN is found to be enhanced in seven positions located around Sgr B2(S), with HOCN to HNCO abundance ratio of   $\sim$ 0.9\%. On the other hand, this  ratio  does not have significant variation in other positions of Sgr B2, ranging from 0.4\% to 0.7\%, indicating that the formation mechanism of both molecules does not vary in most part of Sgr B2. The ratios agree with the gas-grain model well, while the enhancement of HOCN around Sgr B2(S) needs new chemical models to explain it.  
 Based on the intensity ratio of HNCO and HNC$^{18}$O lines, we updated  the isotopic ratio of $^{16}$O/$^{18}$O to be 296 $\pm$ 54 in Sgr B2. 
}

   \authorrunning{S.-Q. Zheng et al. }            %author_head in even pages
   \titlerunning{Spatial distribution of HOCN around Sgr B2}  % title_head in odd pages
   \maketitle

%________________________________________________ sections below
% 
\section{Introduction}           %% first-level sections will be auto-capitalized
\label{sect:intro}

The abundance variation of isomers can give some insights into their formation and destruction \citep{2000ApJ...540L.107H}. The variation can also be used to investigate the environment they exist \citep{1992A&A...256..595S}. Thus, obtaining the relative abundances of isomers is of great importance for getting a deeper understanding of a specific isomer family. The combination of HNCO and  HOCN can be a good choice to study such astro-chemical effect of isomers  \citep{2010ApJ...725.2101Q}.

HNCO is a very abundant molecule with a peptide-like bond. It was first detected in Sagittarius B2 (hereafter Sgr B2) (OH) \citep{1972ApJ...177..619S}. It was regarded as a tracer of dense gas  \citep{1984ApJ...280..608J}. 
It was also proposed to be a tracer of interstellar shocks \citep{2000A&A...361.1079Z}.
%The enhancement of its emission can indicate a slow shock or warm,dense, non-shocked gas  \citep{2017A&A...597A..11K}. 
Many chemical models were used to reproduce the formation of HNCO, such as gas-phase reaction  \citep{1999ApJ...518..699T} and grain-surface reaction  \citep{2008ApJ...682..283G}. 
HOCN is the metastable isomer of HNCO. It was also identified in  Sgr B2   (OH), with an estimated fractional abundance relative to H$_2$ of $1 \times 10^{-11}$ and an abundance ratio relative to HNCO of 0.5\%  \citep{2009ApJ...697..880B}. Then, it was found in other positions towards Sgr B2, including Sgr B2 (M), Sgr B2 (N), and Sgr B2 (S)  \citep{2010A&A...516A.109B}. \cite{2010ApJ...725.2101Q} used gas-grain models to account for the formation of HNCO and its isomers, including HOCN, in different physical environments. The results show that the formation of HNCO involves both gas-phase reaction and grain-surface reaction.

HNCO was observed to have an expanding ring-like emission in Sgr B2 and the peak position is 2\arcsec  ~north of Sgr B2 (M)  \citep{2006NewA...11..594M,2008MNRAS.386..117J}, while HOCN were detected in six positions in the Sgr B2 complex  \citep{2010A&A...516A.109B, 2009ApJ...697..880B}. The enhancement of HNCO can be explained by shocks, which release the molecules in the grain mantles into the gas-phase \citep{2006NewA...11..594M}. There is a fairly constant abundance ratio of ~0.3\%-0.8\% between HOCN and HNCO in the six detected positions \citep{2010A&A...516A.109B}. Therefore, Sgr B2 can be an ideal place to study the relation between HNCO and HOCN. The spatial distribution of HOCN in Sgr B2 and its formation mechanism remain unknown. To further understand the relationship between HNCO and HOCN, mapping observation and extensive surveys in different sources are needed.

%The giant molecular complex Sgr B2, which is located in the central region of the Galaxy at a distance of 8.15 $\pm$ 0.15 kpc \citep{2019ApJ...885..131R}, is a source full of complex organic molecules. There are three massive star-forming regions: Sgr B2(M), (N) and (S). Many complex molecules were first detected in Sgr B2  \citep{2018ApJS..239...17M}. 

 In this paper, we present mapping observations of HNCO and HOCN toward Sgr B2 complex. The outline of this article is presented as followed. In Section 2, we describe the observations and data reduction. In Section 3, we give the mapping result, spatial distribution of HNCO $4_{04}-3_{03}$ and HOCN $4_{04}-3_{03}$ emission, and their column density. Scientific discussions are presented  in Section 4  and a   summary of this paper is given  in Section 5.

% Authors can give a citation as `\citealt{Michel+etal+1992}'.
% You may also use \cite, \citep and \citet for citation, and use Table~1
% or Figure~1 and so forth. Using \ref and \label for cross-references of
% Tables/Figures is a good way in adjusting/adding/removing text, tables or
% figures.

\section{Observations and Data Reduction}
\label{sect:Obs}
We performed point-by-point spectroscopic mapping observations towards Sgr B2 in 2019 May with the IRAM 30m telescope on Pica Veleta, Spain (project 170-18). The observations was performed at 3-mm band. Position-switching mode was adopted.
The broad-band Eight MIxer Receiver and the FFTSs in FTS200 mode were used for the observation. The covered frequency range is 82.3-90 GHz, with  channel spacing  of 0.195 MHz, which  corresponds to the velocity resolution of 0.641 km s$^{-1}$ at 84 GHz. The telescope pointing was checked every $\sim$ 2 hours on 1757-240. The telescope focus was optimized on 1757-240 at the beginning of the observation.
The integration time range from 24 minutes to 98 minutes for different positions, with typical system temperatures  of  $\sim$ 110K,  leading to  1$\sigma$ rms in T$_A^*$ of 4-8 mK derived with  the line free channels.   The antenna temperature ($T_{\rm A}^{\ast}$) was converted to the main beam brightness temperature ($T_{\rm mb}$), using $T_{\rm mb}$=$T_{\rm A}^{\ast}\cdot F_{\rm eff}/B_{\rm eff}$,
where the forward efficiency $F_{\rm eff}$ is 0.95 and beam efficiency $B_{\rm eff}$ is 0.81 for 3 mm band. 

 The observing center is Sgr B2(N) ($\alpha_{J2000}=17^h47^m20^s.0$,$\delta_{J2000}=-28^{\circ}22\arcmin 19.0\arcsec$), with a sampling interval of 30\arcsec. The off position of ($\delta \alpha$, $\delta \beta$)=(-752\arcsec, 342\arcsec) was used \citep{2013A&A...559A..47B}. HNCO $4_{04}-3_{03}$ at 87925.237 MHz, HNCO $4_{14}-3_{13}$ at 87597.330 MHz, HOCN $4_{04}-3_{03}$ at 83900.570 MHz, and HNC$^{18}$O $4_{04}-3_{03}$ at 83191.568 MHz lines within the observing frequency range are used for this study, which are listed in Table \ref{table 1}. The spectroscopic parameters of molecules are taken from CDMS catalog \citep{2005JMoSt.742..215M}.  $4_{04}-3_{03}$ transition is the strongest for both HNCO and HOCN, which is found to be free of confusion from other species.The data processing was conducted using \textbf{GILDAS} software package\footnote{\tt http://www.iram.fr/IRAMFR/GILDAS.}, including CLASS and GREG.

\begin{table*}
\scriptsize
    \begin{center}
      \caption{Transitions of HNCO and HOCN}
      \label{table 1}
      \begin{tabular}{lcccccc}
    \hline
    \hline
      
            &                            &              &      & Sgr B2(N) &     &       Comments        \\
 Molecular &      Transition           & Rest Freq.      &   A   & $E_{u}$ & $\mu^2$S &                  \\
           &                           &  (MHz)          &  $\times$ 10$^{-6}$   &  (K)  & (D$^2$)  &                    \\
\hline        
HOCN    &  4$_{0,4}$-3$_{0,3}$         & 83900.570(0.004) &   42.2    & 10.07   & 55.212  &    No blend      \\     
   
HNC$^{18}$O & 4$_{0,4}$-4$_{0,3},F=3-2 $     & 83191.568(0.030) &     &  9.98   &   7.332    & Partial blend with C$^{13}$H$_2$CHCN  \\
        &  4$_{0,4}$-3$_{0,3},F=5-4 $         & 83191.568(0.030)  &      & 9.98  &   12.546  & Partial blend with C$^{13}$H$_2$CHCN   \\
         &  4$_{0,4}$-4$_{0,3},F=4-3$    & 83191.568(0.030) &    & 9.98   &  9.623  &   Partial blend with C$^{13}$H$_2$CHCN   \\ 
    
HNCO    &  4$_{1,4}$-3$_{1,3}$       & 87597.330(0.008) &  8.04  & 53.79  &  9.254  &  Partial blend with HN$^{13}$CO   \\
           & 4$_{0,4}$-3$_{0,3}$    & 87925.237(0.008) &  8.78 &  10.55 & 9.986  &    No blend   \\ 
          & 4$_{1,3}$-3$_{1,2}$     & 88239.020(0.005)  &  8.22 &53.86 &  9.255  &  Partial blend with HN$^{13}$CO \\ 
                                          
\hline
      \end{tabular}  \\
  \end{center}
  Notes.-: Lines used for mapping. Col. (1): chemical formula; Col. (2): transition quantum numbers \citep{2013A&A...559A..47B}, Col. (3): rest frequency;  Col.(4): emission coefficient A; Col. (5): upper state energy level (K); Col. (6): $\mu^2$ is the dipole moment and S is the line strength; Col. (7) comments.
\end{table*}

%\begin{table}
%\bc
%\begin{minipage}[]{100mm}
%\caption[]{Parameters of 9 objects observed by YFOSC\label{tab1}}\end{minipage}
%\setlength{\tabcolsep}{1pt}
%\small
% \begin{tabular}{ccccccccccccc}
%  \hline\noalign{\smallskip}
%Name& Date& Exposure&$u$& $g$& $r$& $i$& $z$& $Y$& $J$& $H$& $K$& Result\\
%(SDSS J)& &($s$)&&&&&&&&&&\\
%  \hline\noalign{\smallskip}
%075733.86+190403.1&2012-02-26&2700&21.37&20.40&19.45&19.00&18.53&17.88&17.02&16.63&15.68&low S/N\\
%085203.84+020437.7&2012-02-27&6000&21.67&20.99&19.69&19.09&18.66&17.82&17.47&16.66&15.72&low S/N\\
%092740.04-023347.5&2012-02-28&4200&21.01& 20.50&19.55&19.08&18.86&18.42&18.01&17.11&16.42&G star\\
%093345.70-020439.5&2012-02-27&3600&23.77&20.72&19.55&19.47&19.41&19.21&18.66&18.46&17.88&quasar \\
%095023.74+004419.7&2012-02-27&6600&23.97&20.97&19.75&19.48&19.36&18.83&18.40&17.76&17.25&quasar\\
%113816.85+045023.6&2012-02-27&6000&21.26&20.70&19.67&19.27&19.06&18.30&17.96&16.90&16.46&quasar\\
%120312.63-001118.8&2012-02-28&5400&25.38&22.34&20.29&19.14&18.95&18.32&18.00&17.19&16.64&quasar\\
%125052.11+074919.6&2012-02-26&5400&23.56&20.08&18.75&18.63&18.43&18.14&17.35&16.81&16.15&quasar \\
%145115.89+015843.3&2012-02-27&4800&23.72&20.42&19.30&19.23&19.09&18.61&18.02&17.56&16.92&quasar \\
%  \noalign{\smallskip}\hline
%\end{tabular}
%\ec
%% place \tablecomments and \tablerefs below \end{center| and \end{center}:
%% you may leave the table-width parameter to editors or set to your actual size
%\tablecomments{0.86\textwidth}{The SDSS $ugriz$ magnitudes are given in AB system 
%and the UKIDSS $YJHK$ magnitudes are given in Vega system.}
%\end{table}

\section{Results}

\begin{figure*}
\centering
\includegraphics[width=0.47\textwidth]{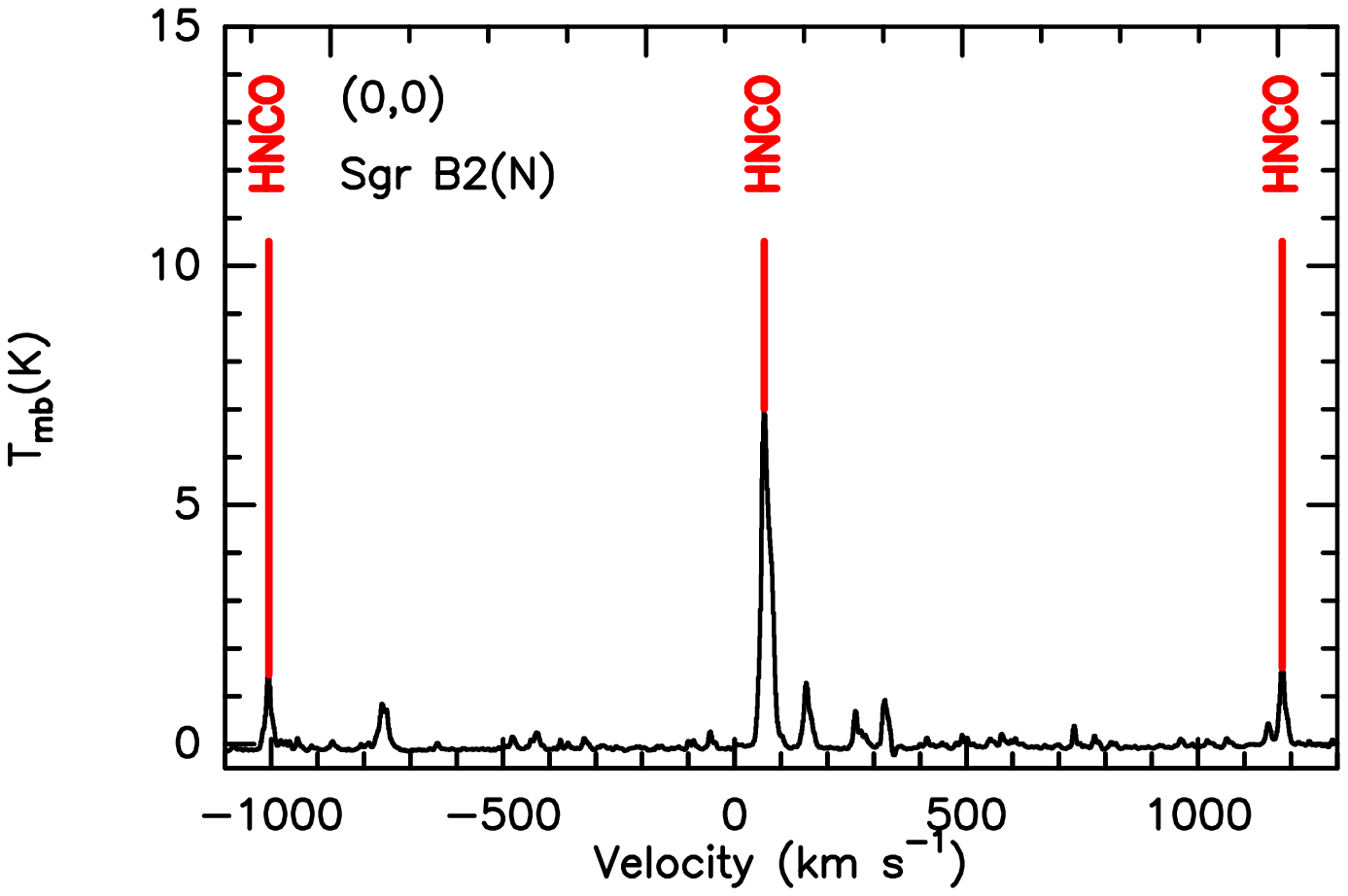}
\includegraphics[width=0.47\textwidth]{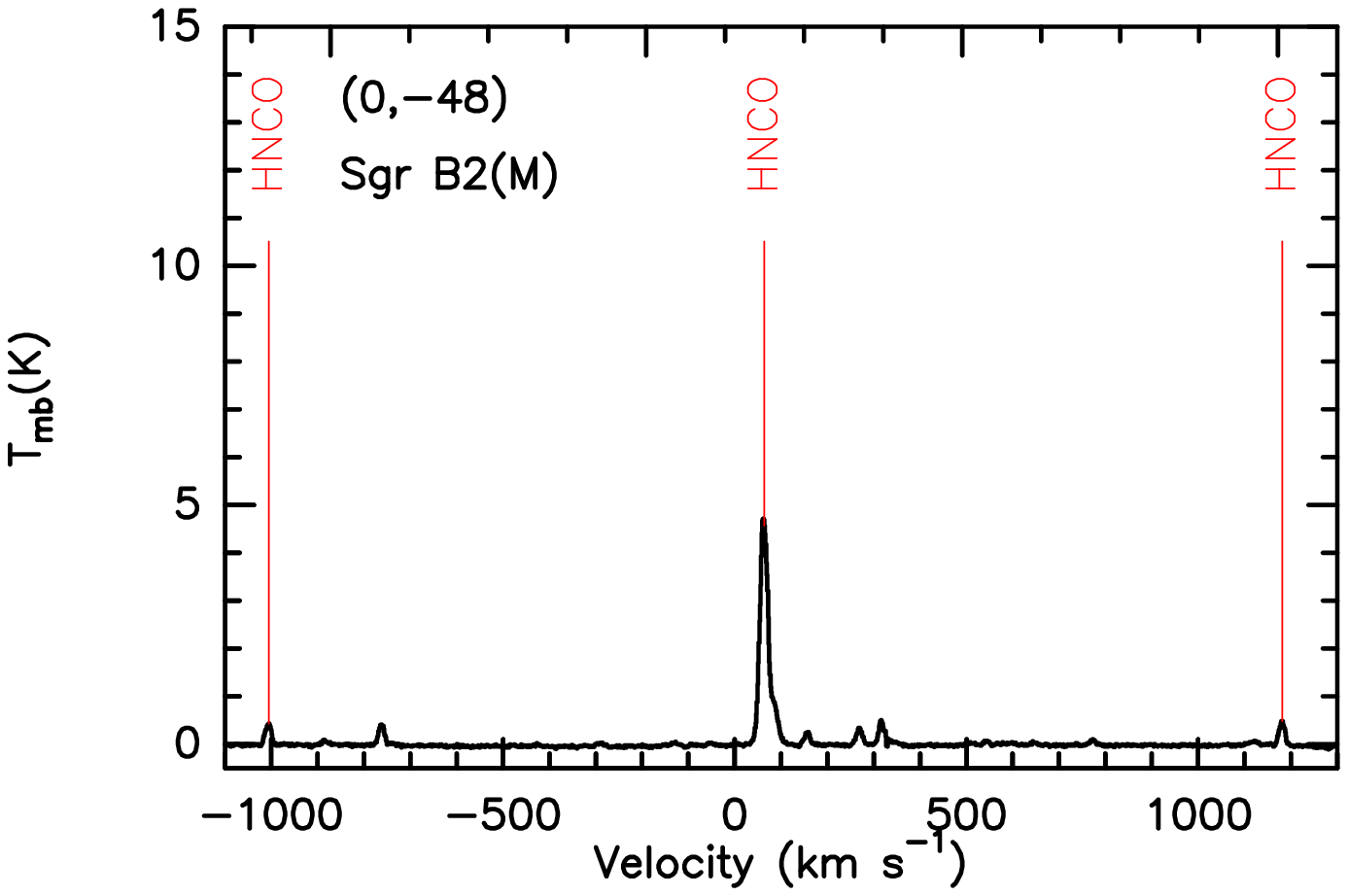}
\includegraphics[width=0.47\textwidth]{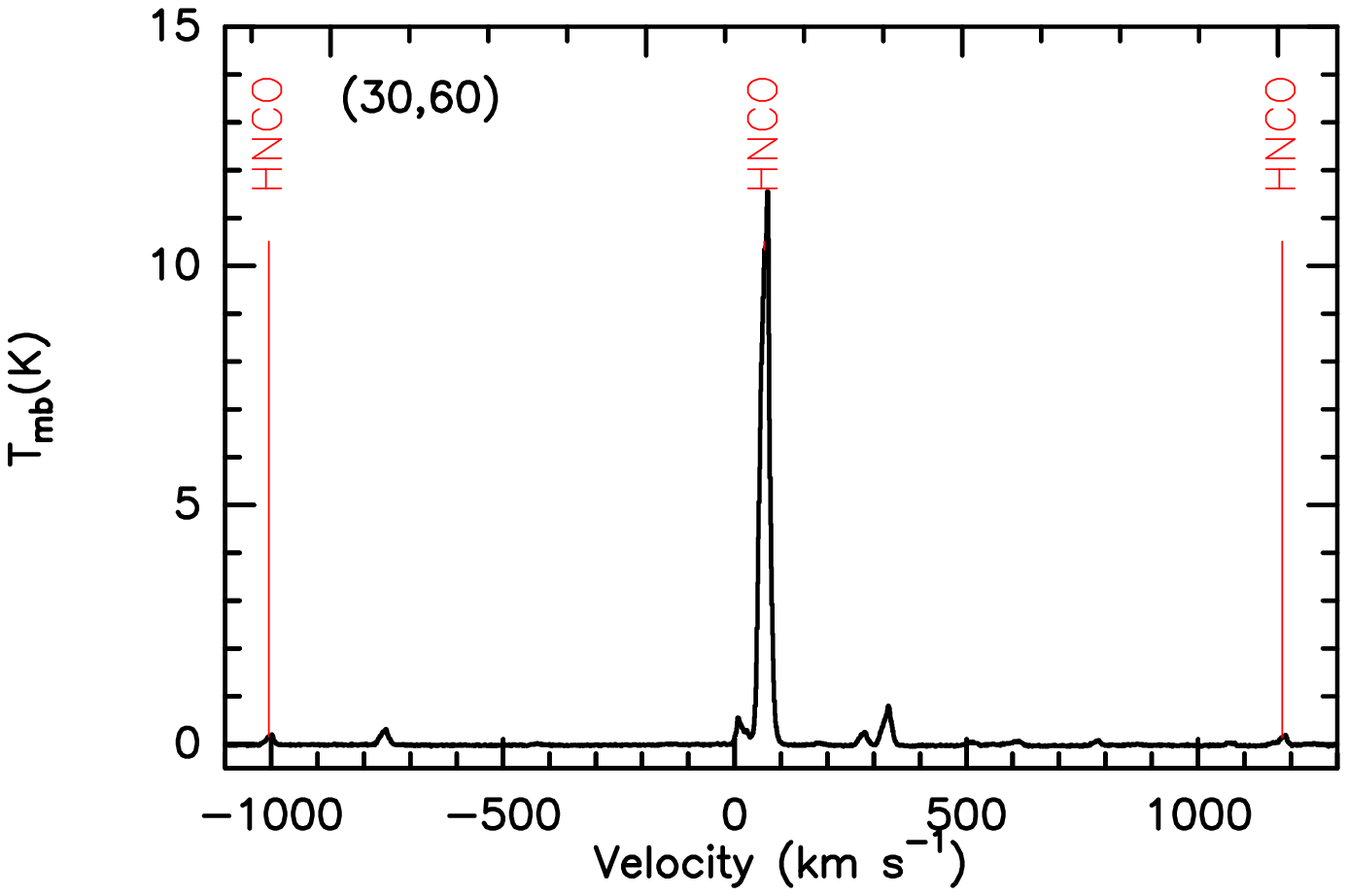}
\includegraphics[width=0.47\textwidth]{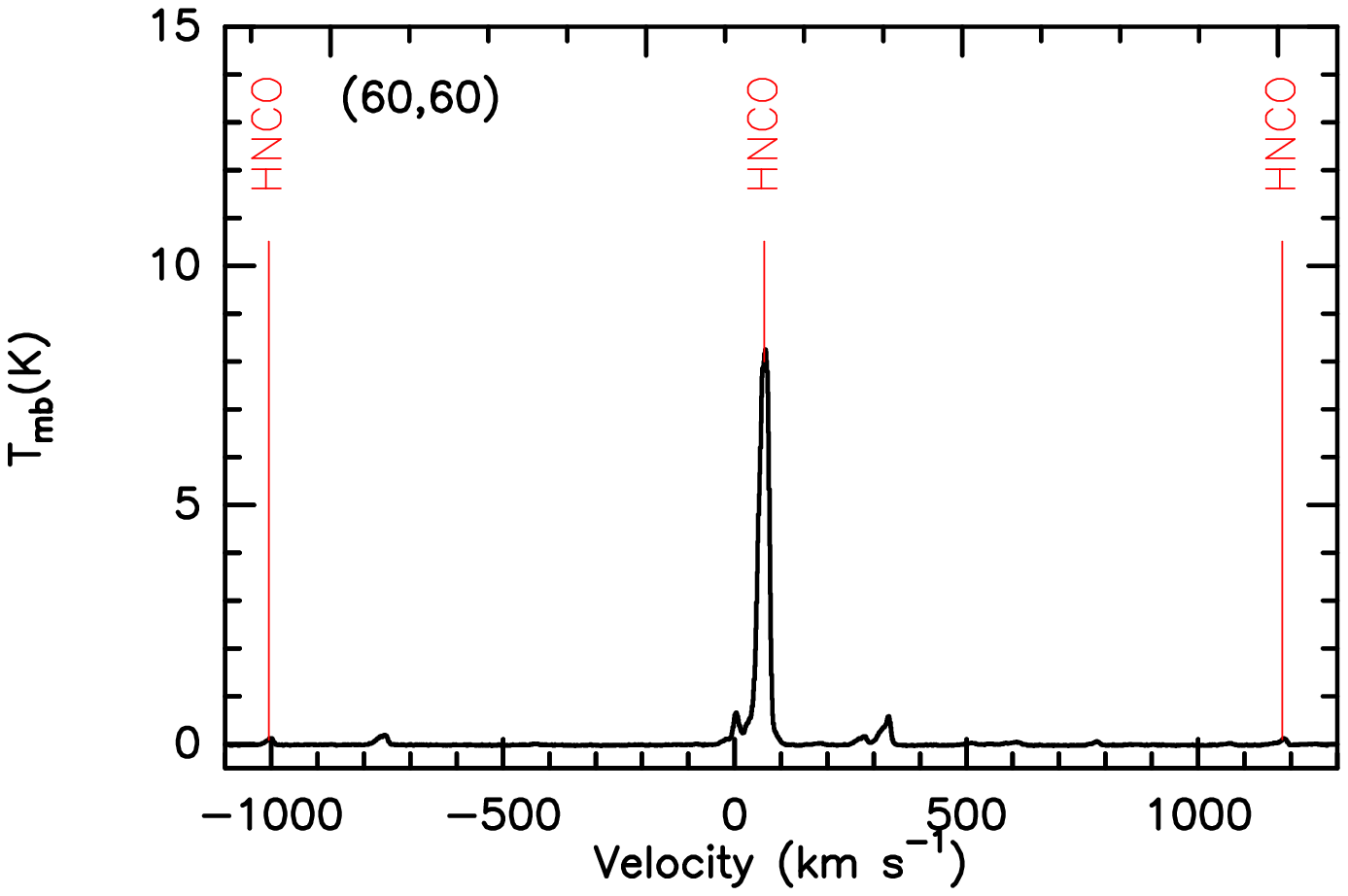}
\caption{The spectra of HNCO $4_{04}-3_{03}$ toward positions (0,0), (0,-48), (30,60) and (60, 60), corresponding to Sgr B2(N), Sgr B2(M), the strongest emission and other common positions. The red lines mark the place of different transitions.}
\label{fig 1}
\end{figure*}

\begin{figure*}
\centering
\includegraphics[width=0.47\textwidth]{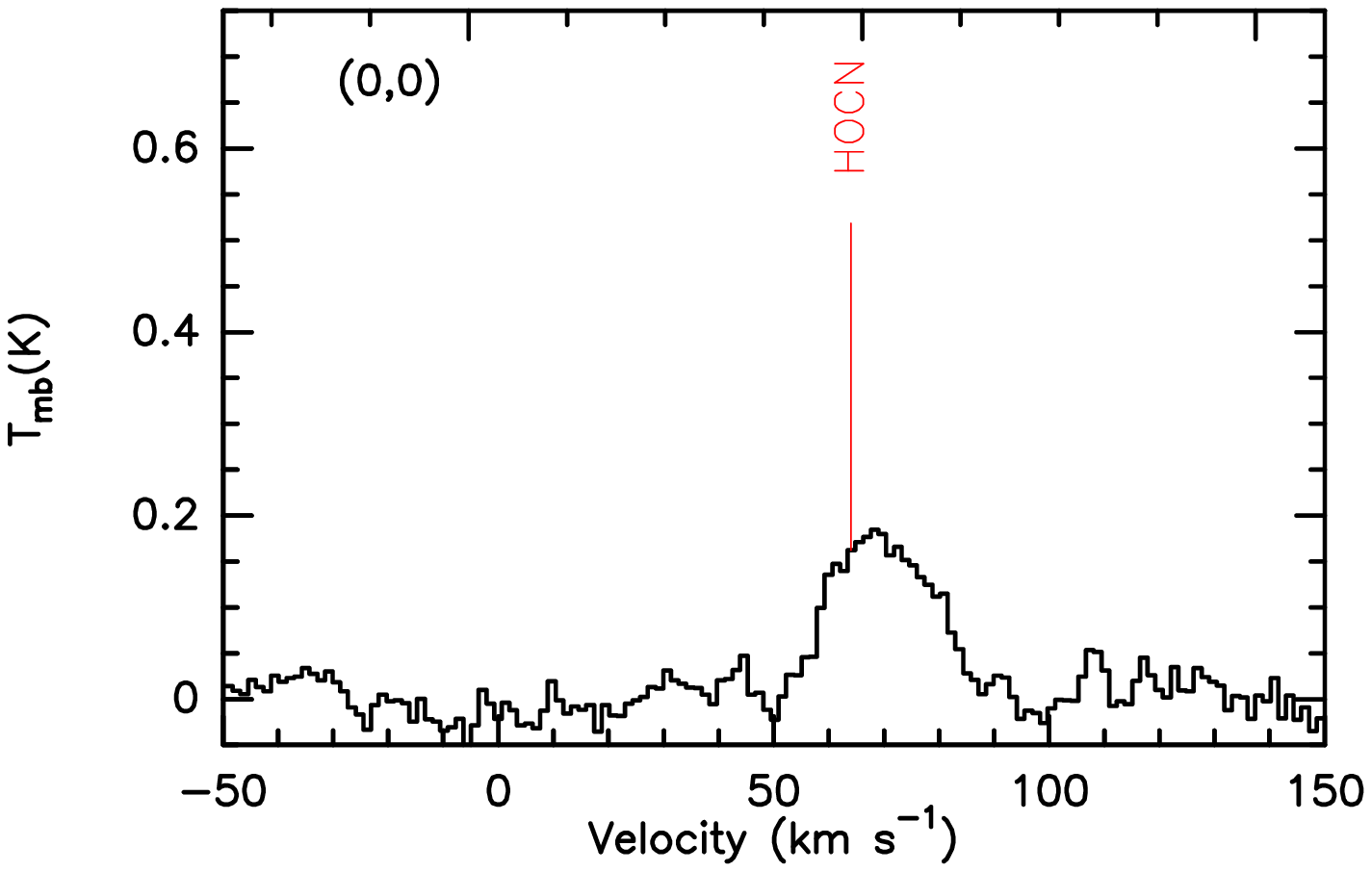}
\includegraphics[width=0.47\textwidth]{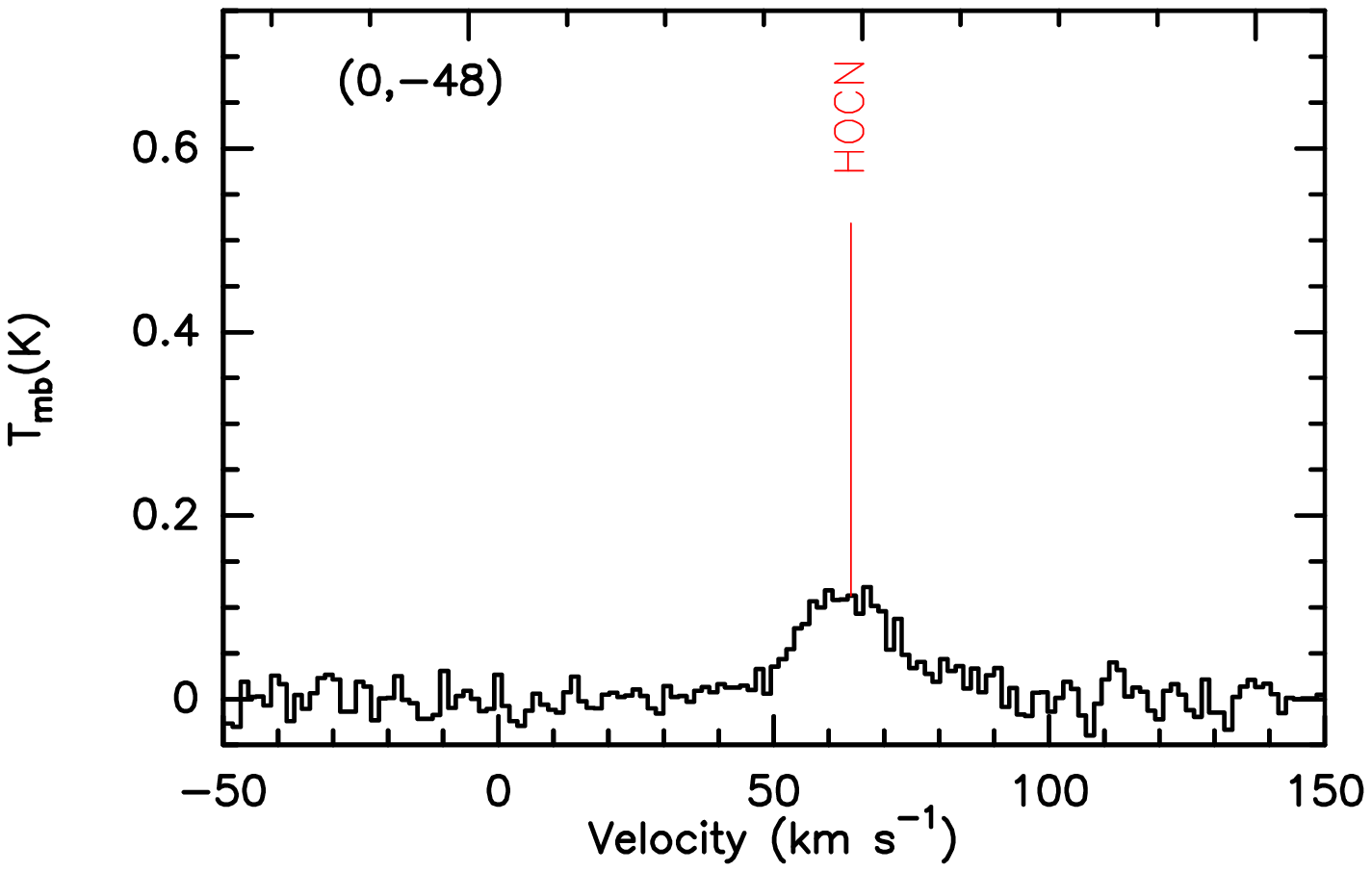}
\includegraphics[width=0.47\textwidth]{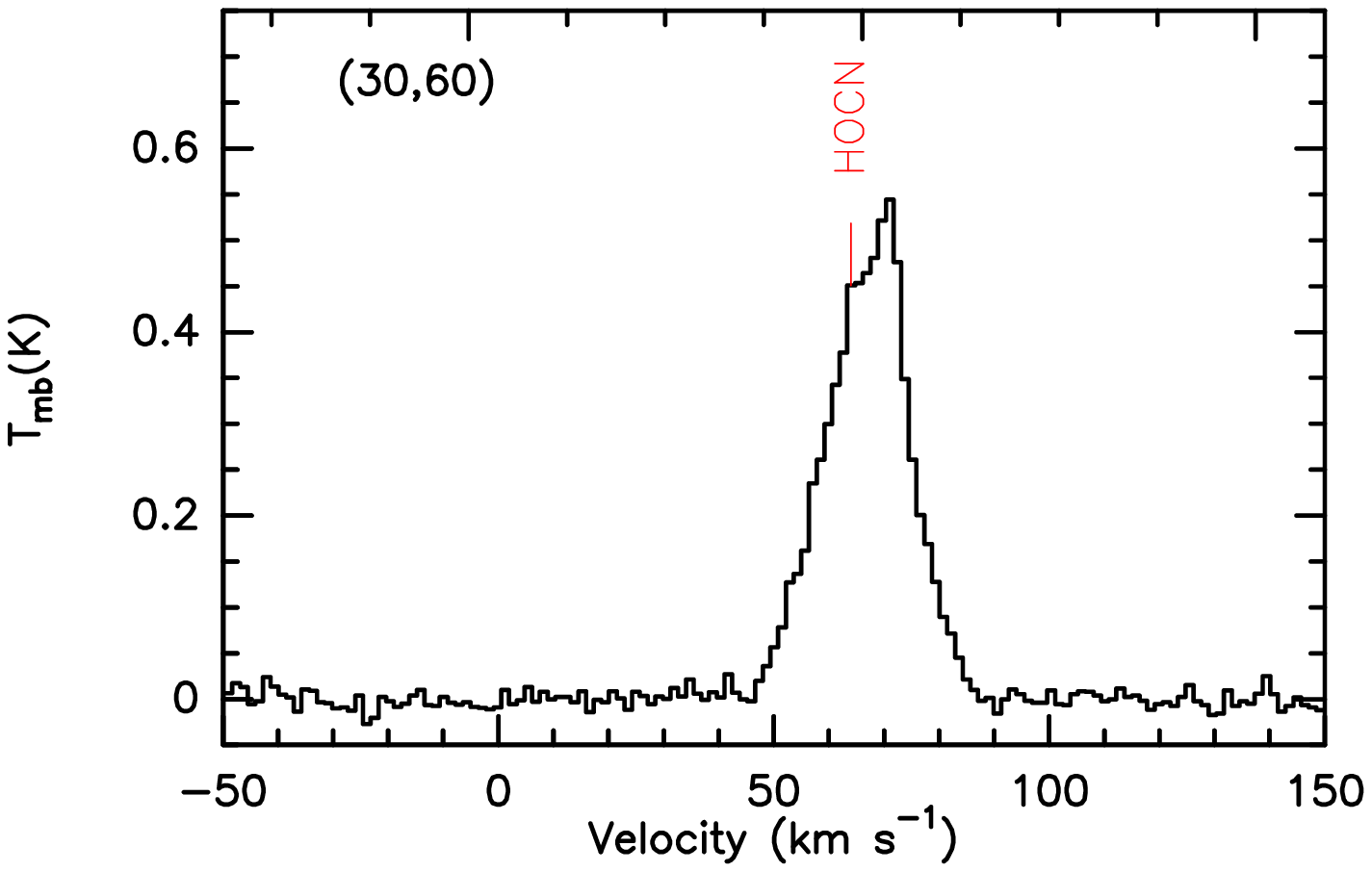}
\includegraphics[width=0.47\textwidth]{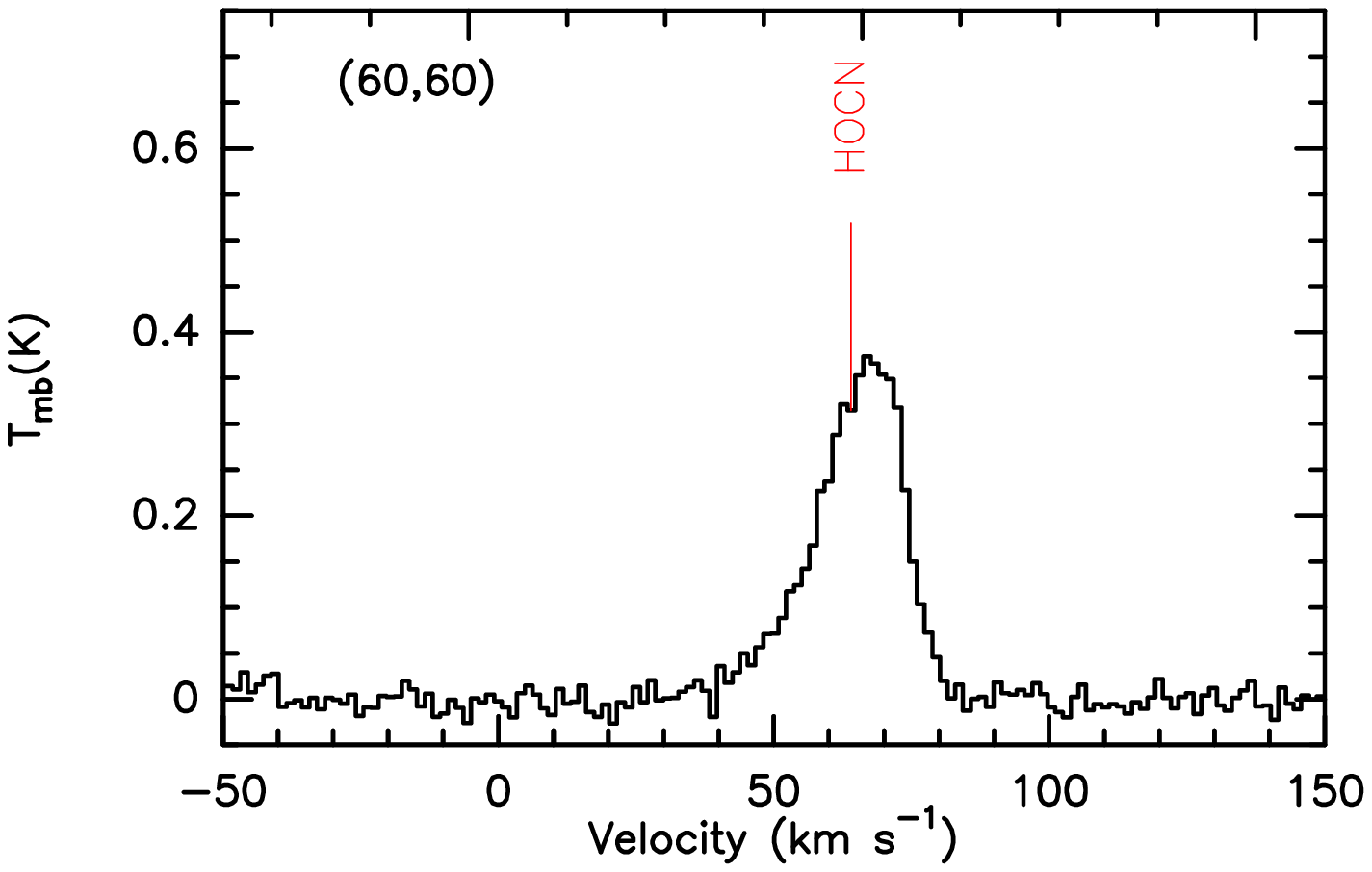}
\caption{HOCN $4_{04}-3_{03}$ spectra at the rest frequency of 83900.570 MHz toward 4 positions of Sgr B2. The red lines mark the transition of HOCN.}
\label{fig 2}
\end{figure*}

\begin{figure*}
\includegraphics[width=0.47\textwidth]{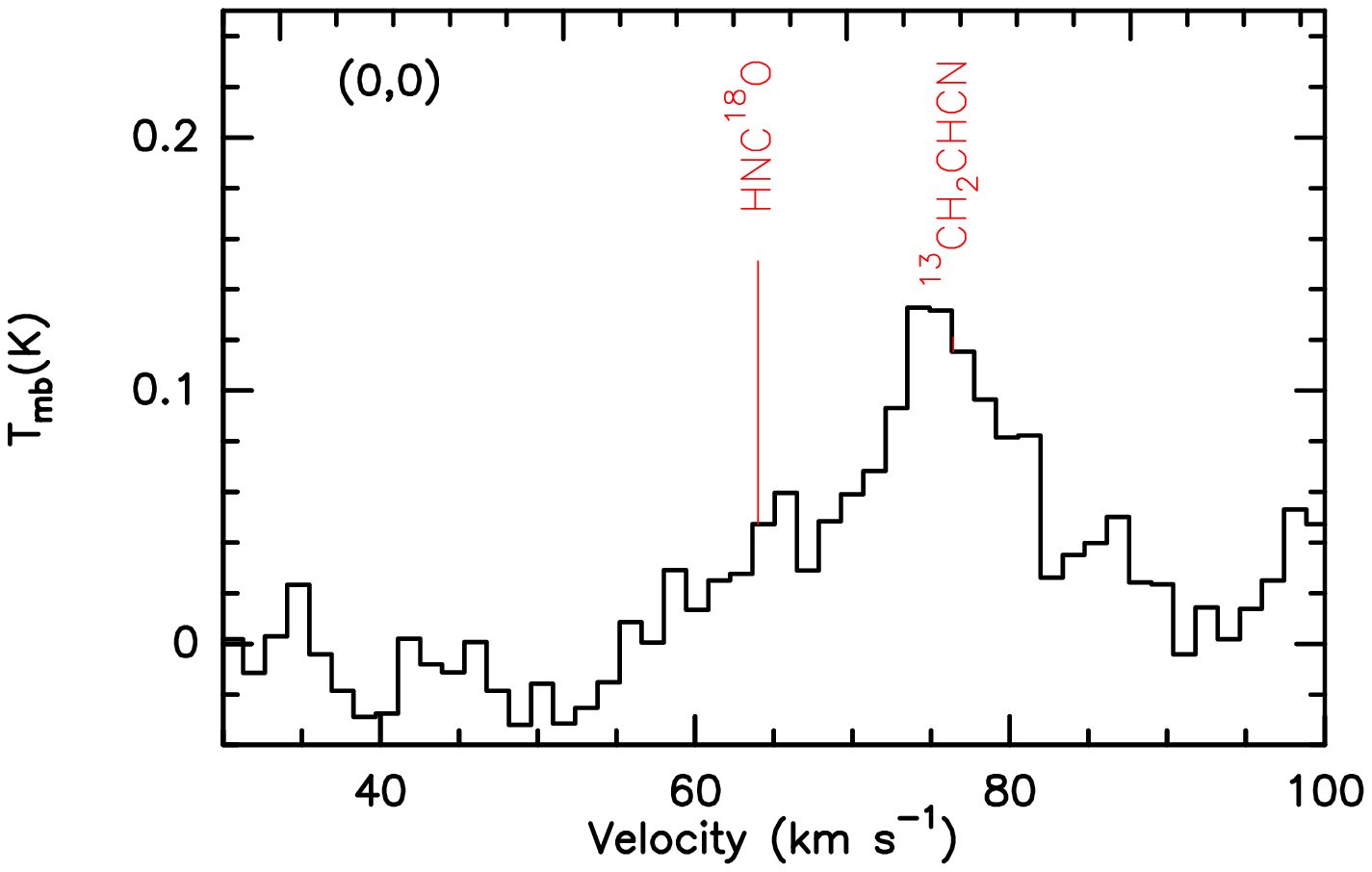}
\includegraphics[width=0.47\textwidth]{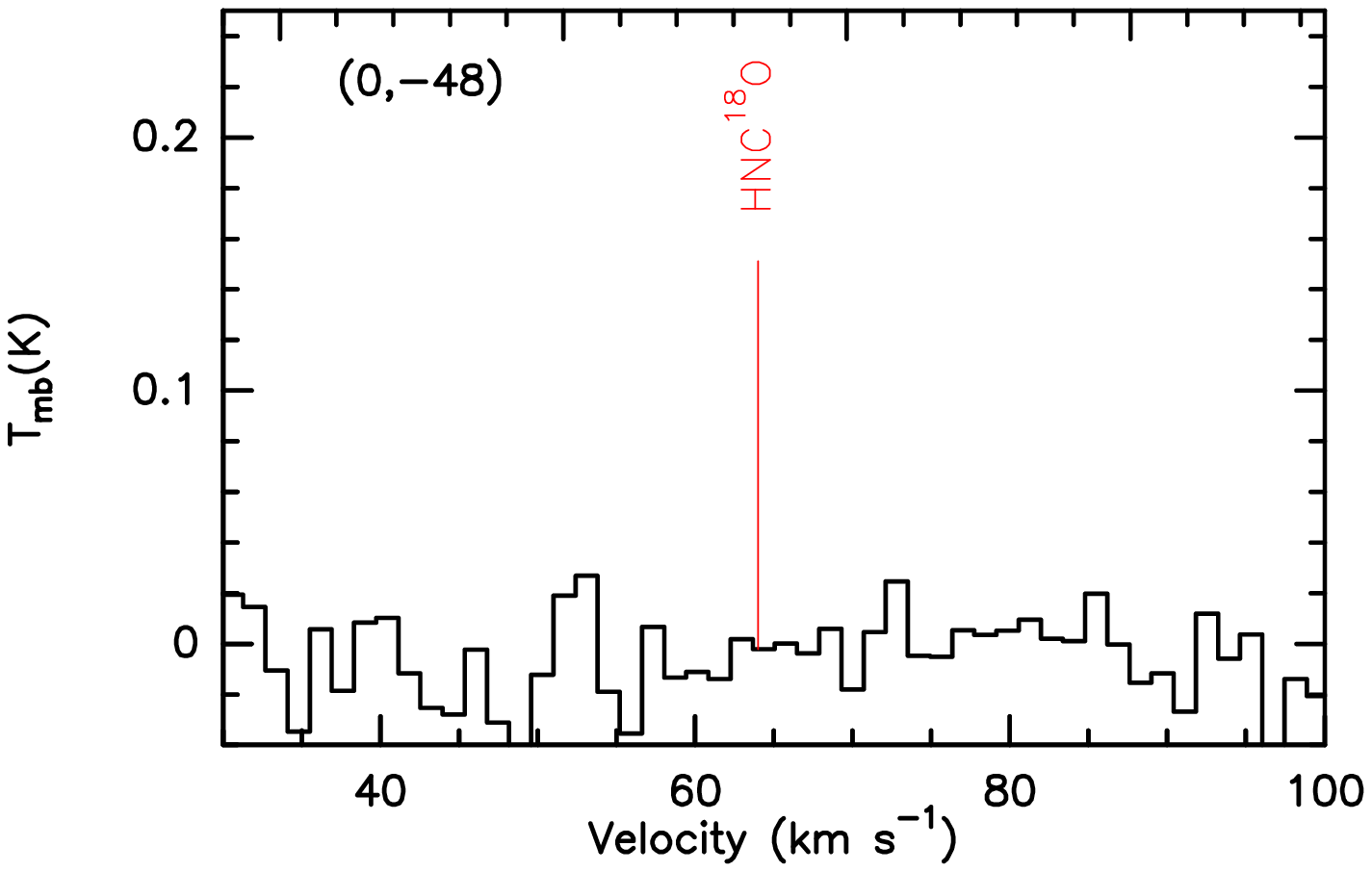}
\includegraphics[width=0.47\textwidth]{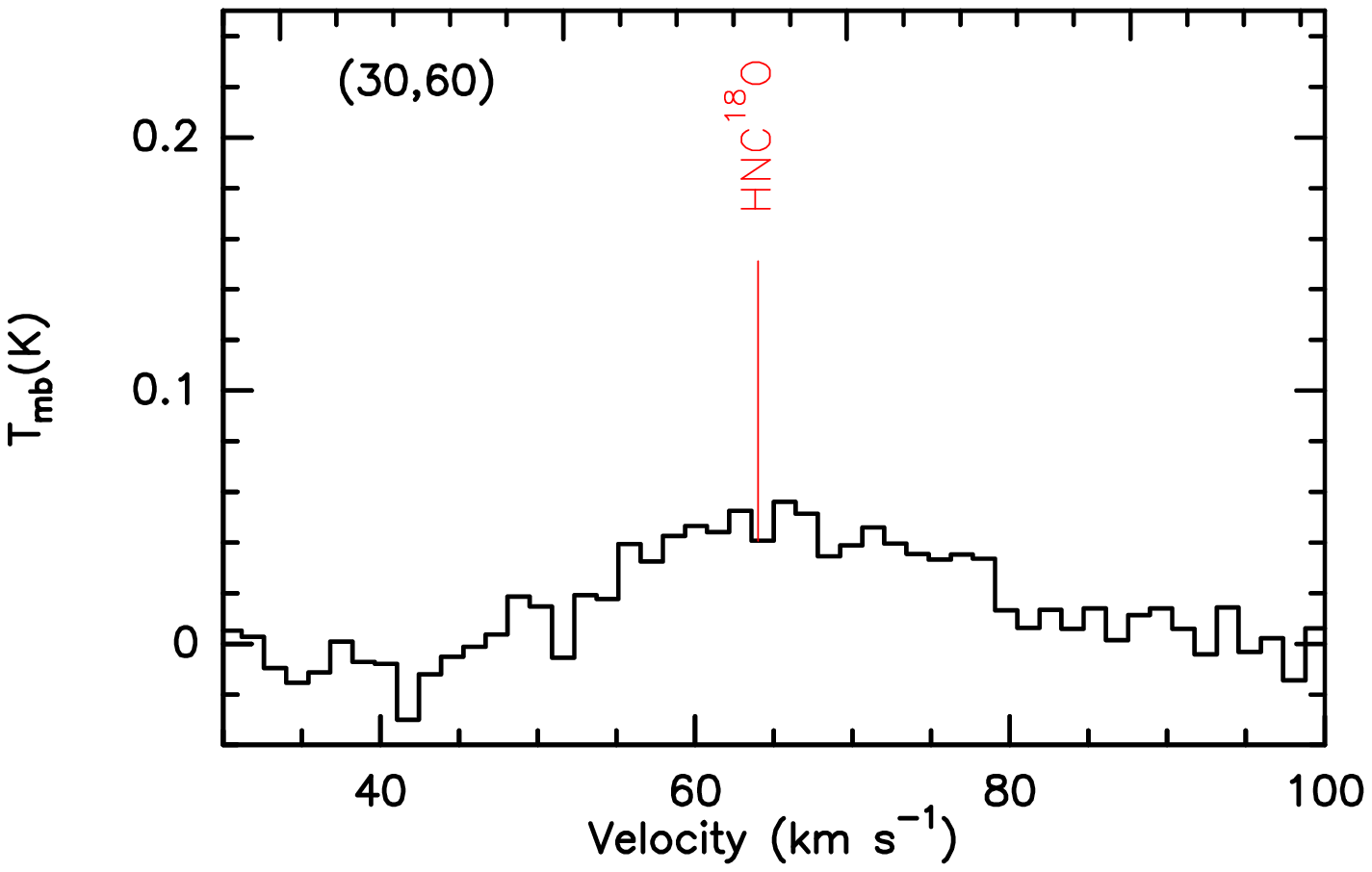}
\includegraphics[width=0.47\textwidth]{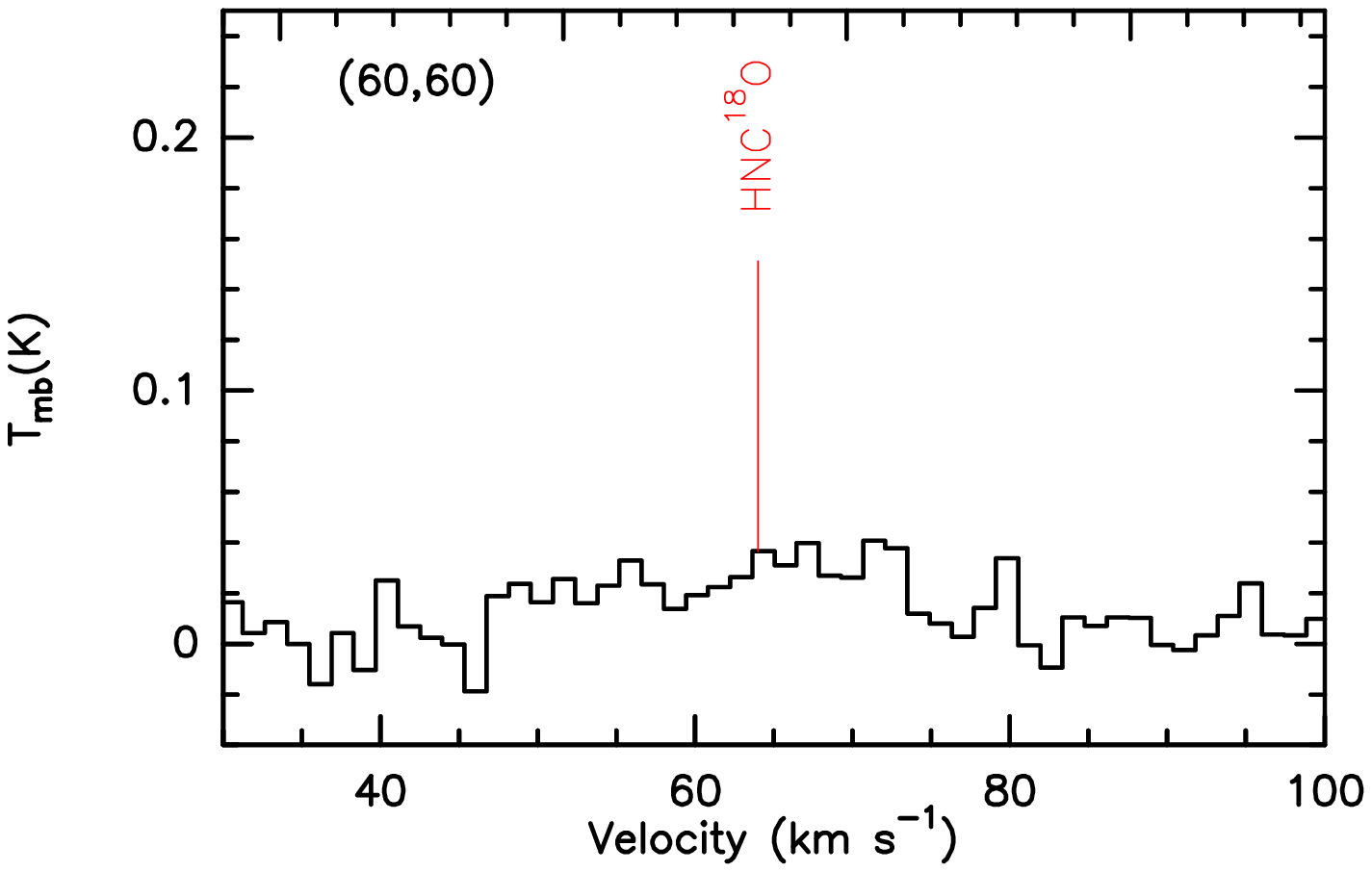}
\caption{HNC$^{18}$O $4_{04}-3_{03}$ spectra at the rest frequency of 83191.568 MHz toward 4 positions of Sgr B2. The red lines correspond to the transition of HNC$^{18}$O and the transition of $^{13}$CH$_2$CHCN at 83188.122 MHz.}
\label{fig 3}
\end{figure*}

%The emission of HNCO was detected above 5 $\sigma$ levels in all the points, and the emission of HOCN was detected in 57 out of the 63 points above 5 $\sigma$ level. The spectra of HNCO$4_{04}-3_{03}$ and HOCN $4_{04}-3_{03}$ transitions towards several points in Sgr B2 are presented in Figure \ref{fig 1} and Figure \ref{fig 2}, which  shows that the emission of HNCO $4_{04}-3_{03}$  is about  40 times of HOCN $4_{04}-3_{03}$. The emission of HNC$^{18}$O was detected toward 33 points, while HNCO$4_{14}-3_{13}$ was detected toward 58 points above 3 $\sigma$ level (see Figure \ref{fig 3} and Figure \ref{fig 1}). 
We have observed 63 positions toward Sgr B2 to obtain the spatial properties of HNCO and HOCN there. HNCO $4_{04}-3_{03}$ emission was observed to be strong toward all the positions. The emission of HOCN $4_{04}-3_{03}$ was detected to be above 5 $\sigma$ level in 57 positions. The emission of HNC$^{18}$O $4_{04}-3_{03}$ was detected to be above 3 $\sigma$ level in 33 positions, while HNCO $4_{14}-3_{13}$ was detected toward 58 positions above 3 $\sigma$ level. 
The spectra of the lines mentioned above are presented in Figure \ref{fig 1}, \ref{fig 2}, \ref{fig 3} towards four positions, which are Sgr B2(N), Sgr B2(M), (30, 60), and (60, 60), as examples. The strongest HOCN $4_{04}-3_{03}$ emission comes from (30, 60). From those spectra, we can find that the emission of HNCO $4_{04}-3_{03}$  is about  40 times of HOCN $4_{04}-3_{03}$.
No significant line blending for HNC$^{18}$O $4_{04}-3_{03}$ at 83191.568 MHz is found, except for that toward Sgr B2(N), which is strongly blended with  $^{13}$CH$_2$CHCN transitions.    The line profile can  not  be  fitted with one simple Gaussian profile for most of the positions,  which implies that there are multiple velocity components. To obtain a more accurate value of intensity, we integrate the spectra directly to get the intensity in the later analysis.

\subsection{Spatial distribution of HNCO and HOCN emission}
\begin{figure*}
\centering
\includegraphics[width=0.4\textwidth]{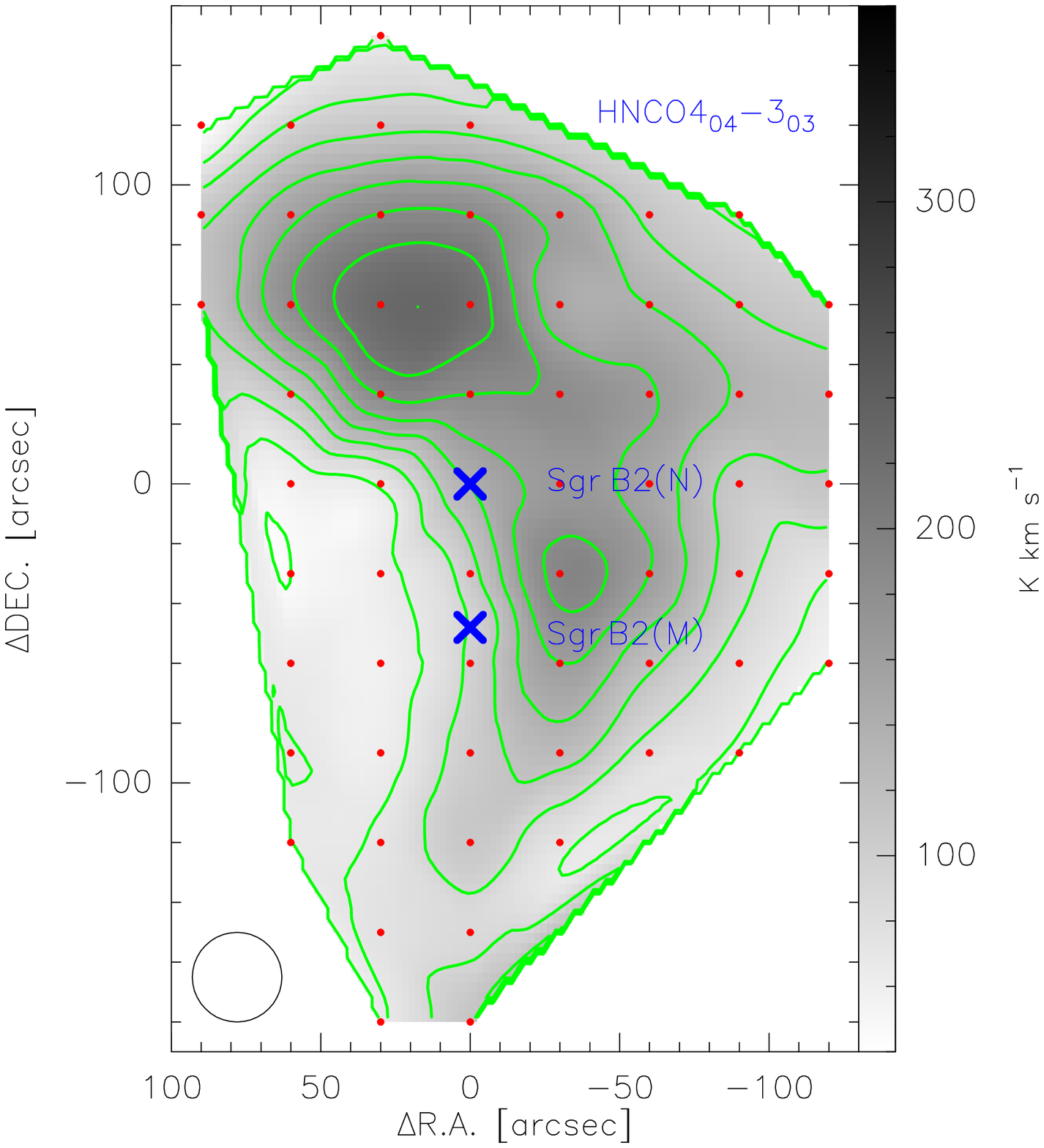}
\includegraphics[width=0.4\textwidth]{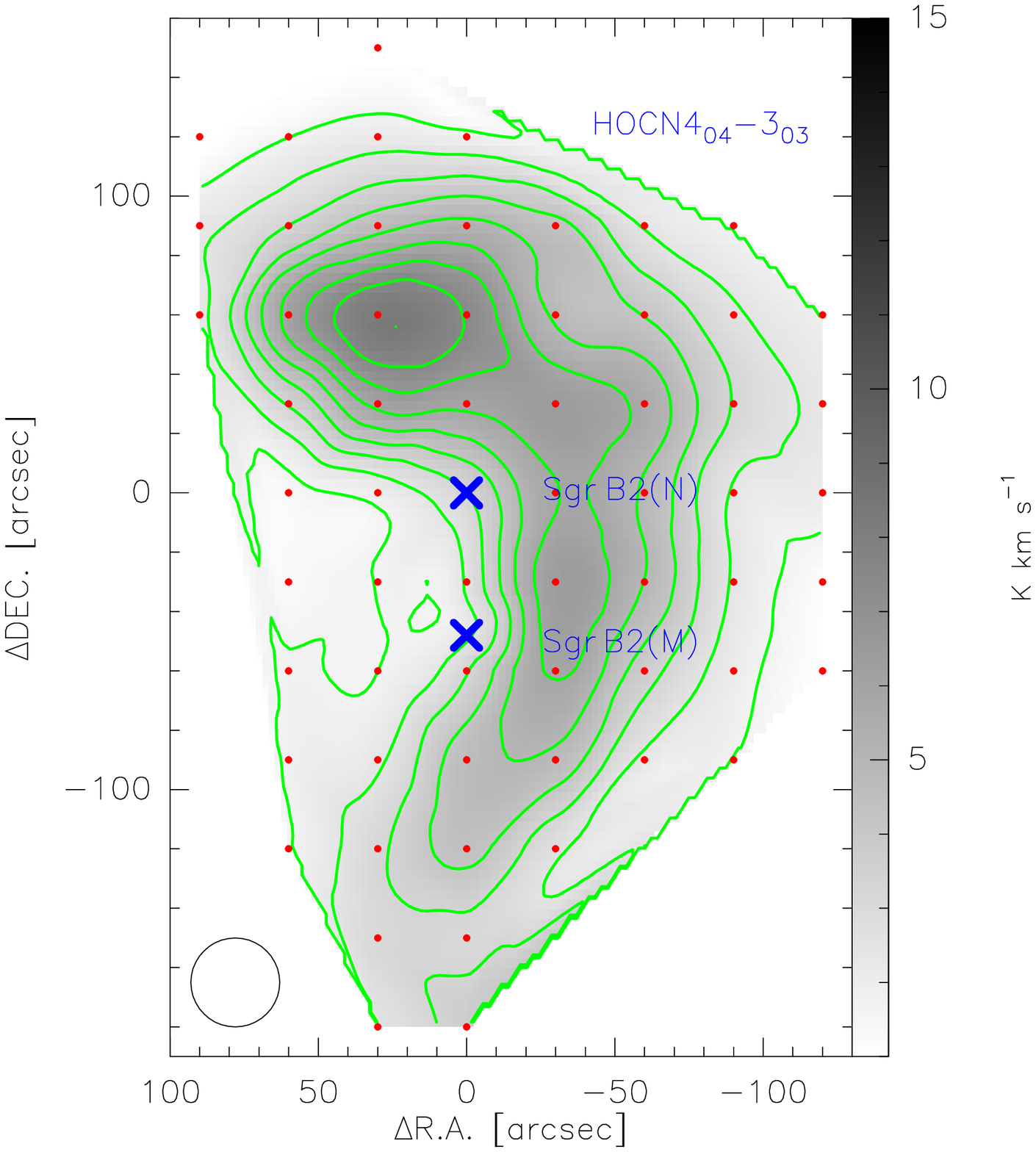}
\includegraphics[width=0.4\textwidth]{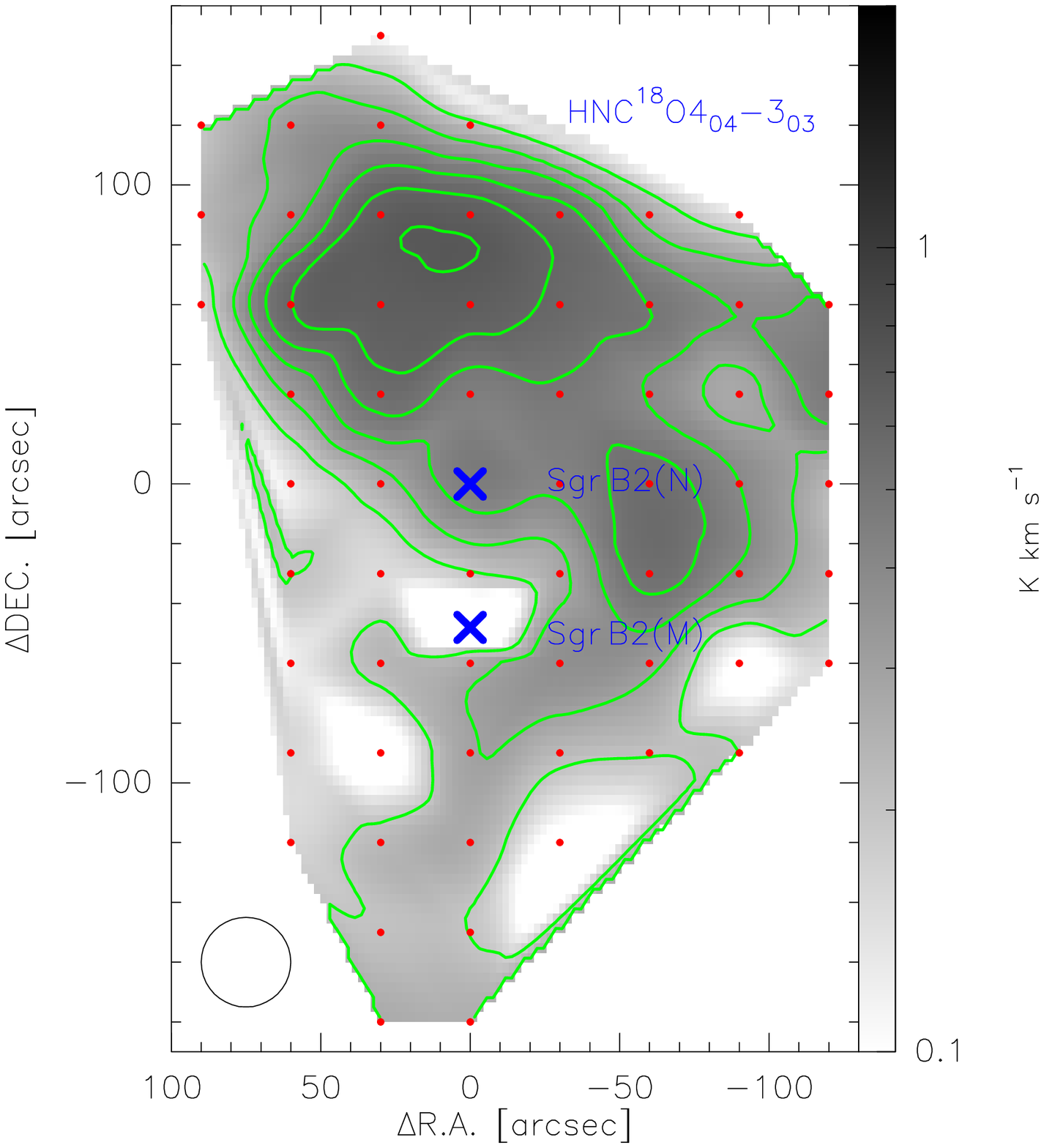}
\includegraphics[width=0.4\textwidth]{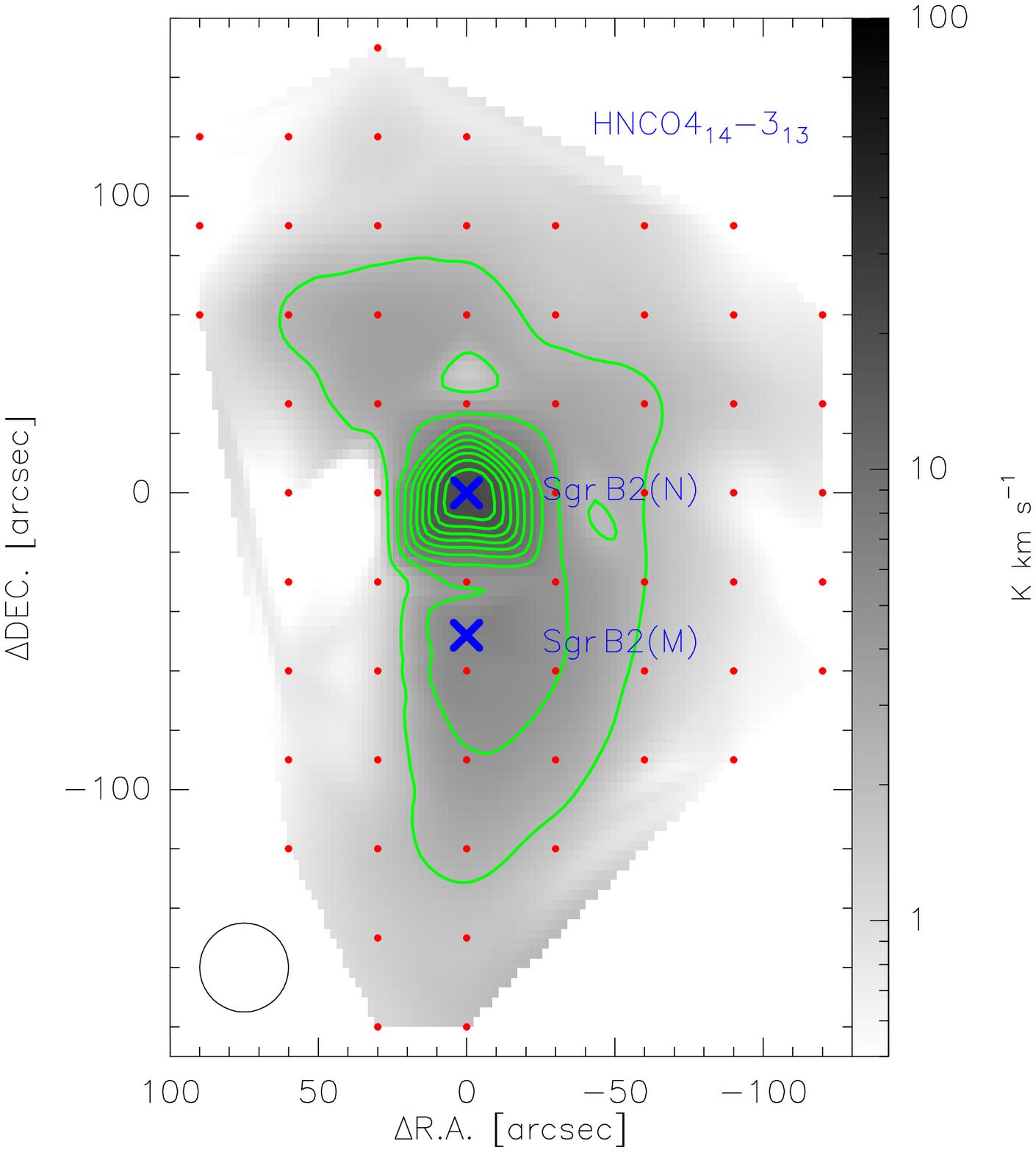}
\caption{The gray-scale and contours show the integrated emission from HNCO $4_{04}-3_{03}$, HOCN $4_{04}-3_{03}$, HNC$^{18}$O $4_{04}-3_{03}$,  and  HNCO $4_{14}-3_{13}$  in Sgr B2. The contour levels of HNCO $4_{04}-3_{03}$, HOCN $4_{04}-3_{03}$ and HNC$^{18}$O $4_{04}-3_{03}$  range from 20\% $\sim$ 90\% with the step of 10\%. To show the structure of HNCO $4_{14}-3_{13}$, the contour levels range from 10\% $\sim$ 90\% with the step of 10\%. The peak values for each panel are 226.3 K km s$^{-1}$, 8.1 K km s$^{-1}$, 0.66 K km s$^{-1}$ and 21.8 K km s$^{-1}$ successively. $``\times"$ indicates Sgr B2(N) and Sgr B2(M). The beams are shown in the left corner of each panel.}
\label{fig 4}
\end{figure*}

The velocity integrated intensity maps of $4_{04}-3_{03}$, the strongest line for HNCO and HOCN, are presented in Figure \ref{fig 4}. In the figure, the positions of Sgr B2 (N) and (M) were marked as $``\times"$. The contour levels range from 20\% $\sim$ 90\% with the step of 10\%. These two maps resemble each other very well. According to the maps, the emission of HNCO $4_{04}-3_{03}$ and HOCN $4_{04}-3_{03}$ are extended with an expanding ring like morphology and peaked at the north of Sgr B2, avoiding the hot cores Sgr B2 (M) and (N), which is similar to the map of HNCO $5_{05}-4_{04}$ emission  \citep{2008MNRAS.386..117J}.

%Among the 63 points we observed, all of the HNCO $4_{04}-3_{03}$ transition are strong and obvious, while HOCN $4_{04}-3_{03}$ is easily distinguished only in 57 points  due to its lower abundance. 

The velocity  integrated intensity maps of HNC$^{18}$O $4_{04}-3_{03}$ and HNCO $4_{14}-3_{13}$ are also presented in Figure \ref{fig 4}. 
%As there is contamination of $^{13}$CH$_2$CHCN, the integrated intensity of HNC$^{18}$O should be calculated after subtracting the  $^{13}$CH$_2$CHCN contribution. 
The profiles of HNC$^{18}$O $4_{04}-3_{03}$ are  unblended in some positions, in which the intensity of HNC$^{18}$O $4_{04}-3_{03}$ can be simply obtained. For the other positions where the profile of HNC$^{18}$O and $^{13}$CH$_2$CHCN emission can not be distinguished, the intensity of $^{13}$CH$_{2}$CHCN is needed. As the emission of $^{13}$CH$_{2}$CHCN is weak in other positions except for (0,0),   the intensity of $^{13}$CH$_{2}$CHCN can be estimated with the line emission of  CH$_{2}$CHCN. Assuming a constant ratio of $^{13}$CH$_{2}$CHCN and CH$_{2}$CHCN in different positions of Sgr B2,  with the ratio between $^{13}$CH$_{2}$CHCN and CH$_{2}$CHCN lines at position (0,0), the intensity of $^{13}$CH$_{2}$CHCN blended with HNC$^{18}$O $4_{04}-3_{03}$   in other positions can be obtained, with a fraction of $\sim 0.3\%$ to HNC$^{18}$O $4_{04}-3_{03}$. 
The emission of HNC$^{18}$O is similar to that of HNCO, with an extended spatial distribution. 
The differences between the maps of isotopes may indicate that the emission of HNCO $4_{04}-3_{03}$ is optically thick in some positions. The emission of HNCO $4_{14}-3_{13}$ is also extended, with two peaks located in the hot cores, Sgr B2 (N) and (M). 
Considering the larger upper energy  ($E_u$=53.78 K) of this transition, the two  peaks should be mainly caused by the high temperature of the gas there. 

\subsection{Isotopic ratio $^{16}$O/$^{18}$O}
\begin{figure*}
\centering
\includegraphics[width=0.47\textwidth]{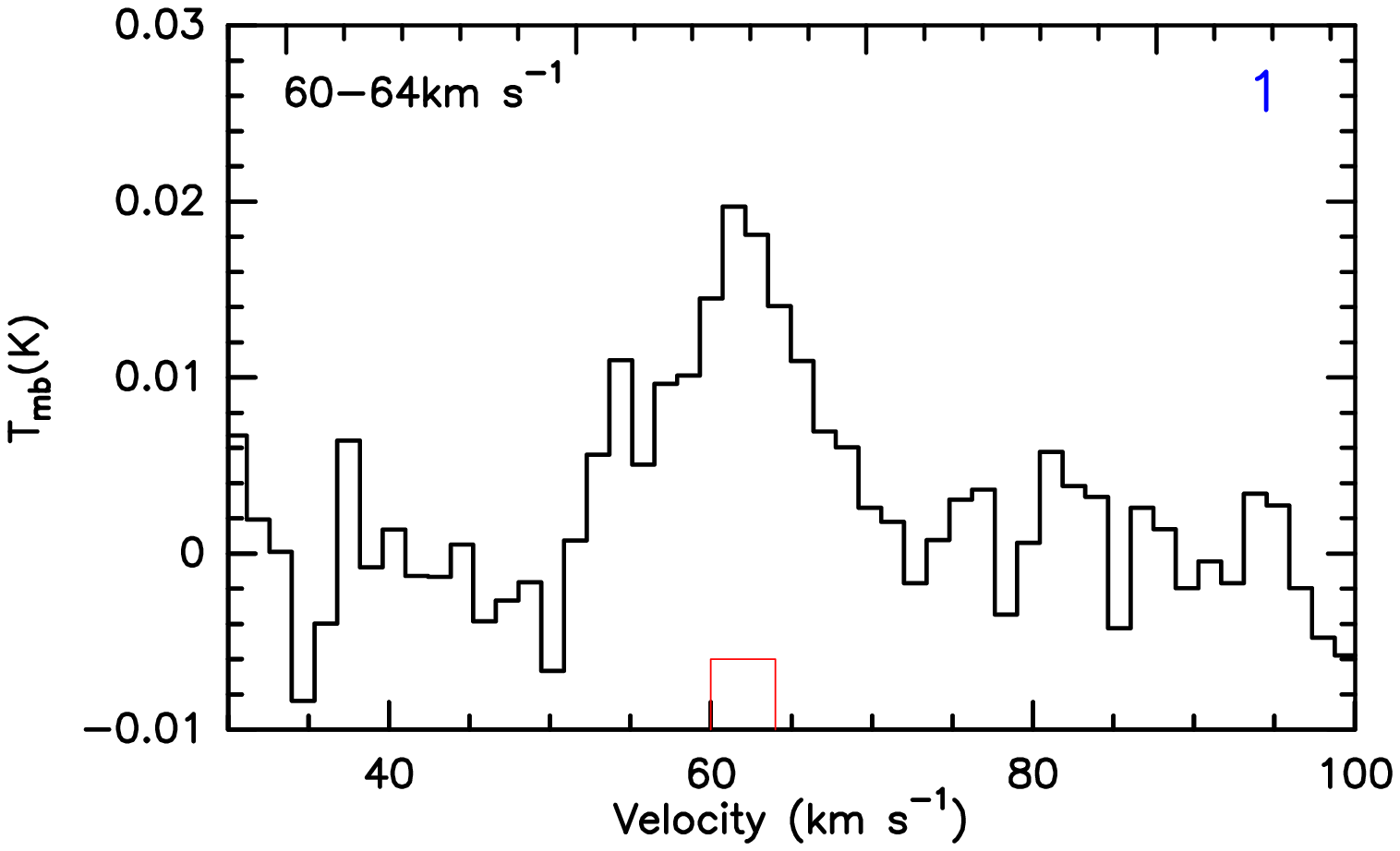}
\includegraphics[width=0.47\textwidth]{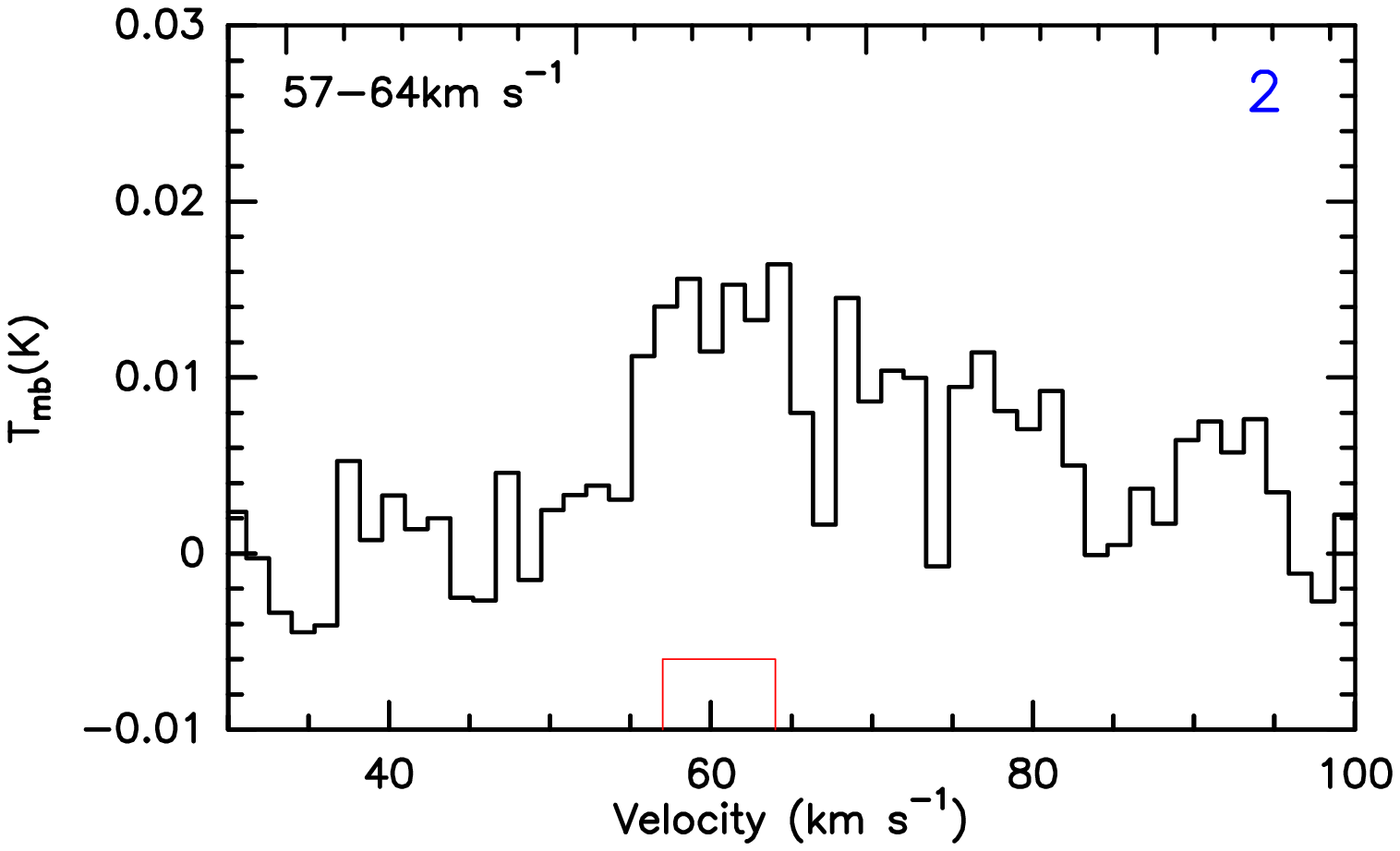}
\includegraphics[width=0.47\textwidth]{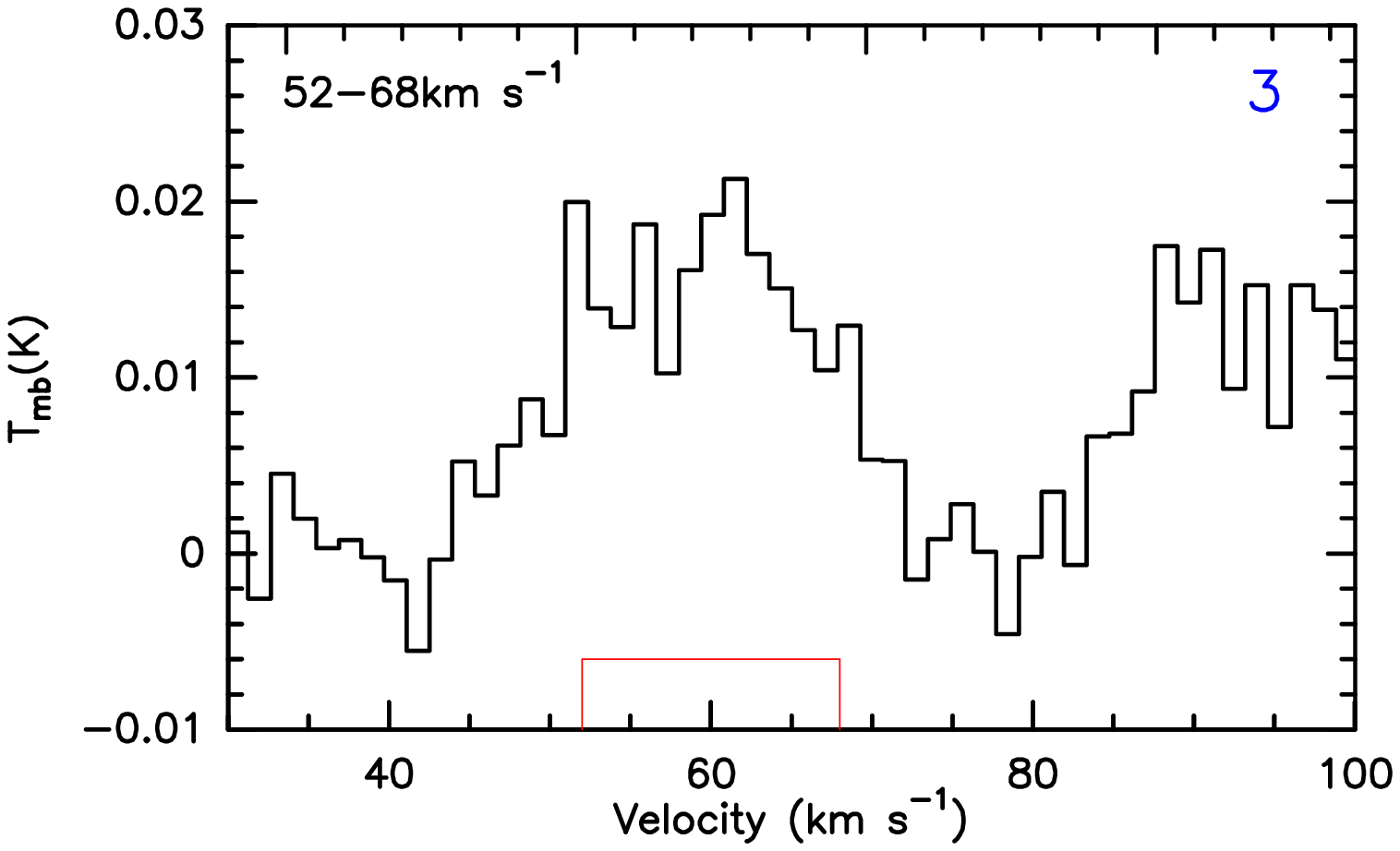}
\includegraphics[width=0.47\textwidth]{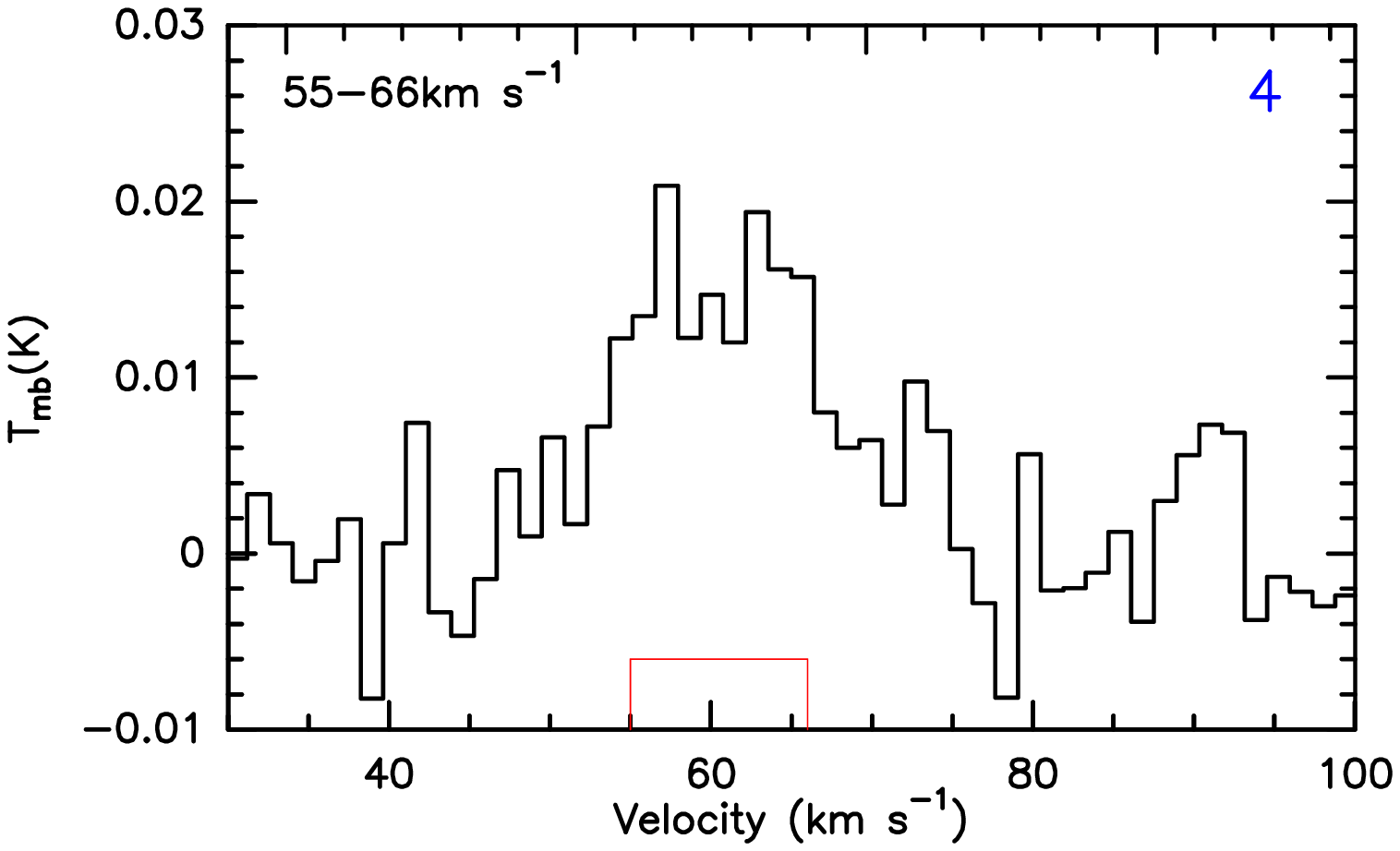}
\includegraphics[width=0.47\textwidth]{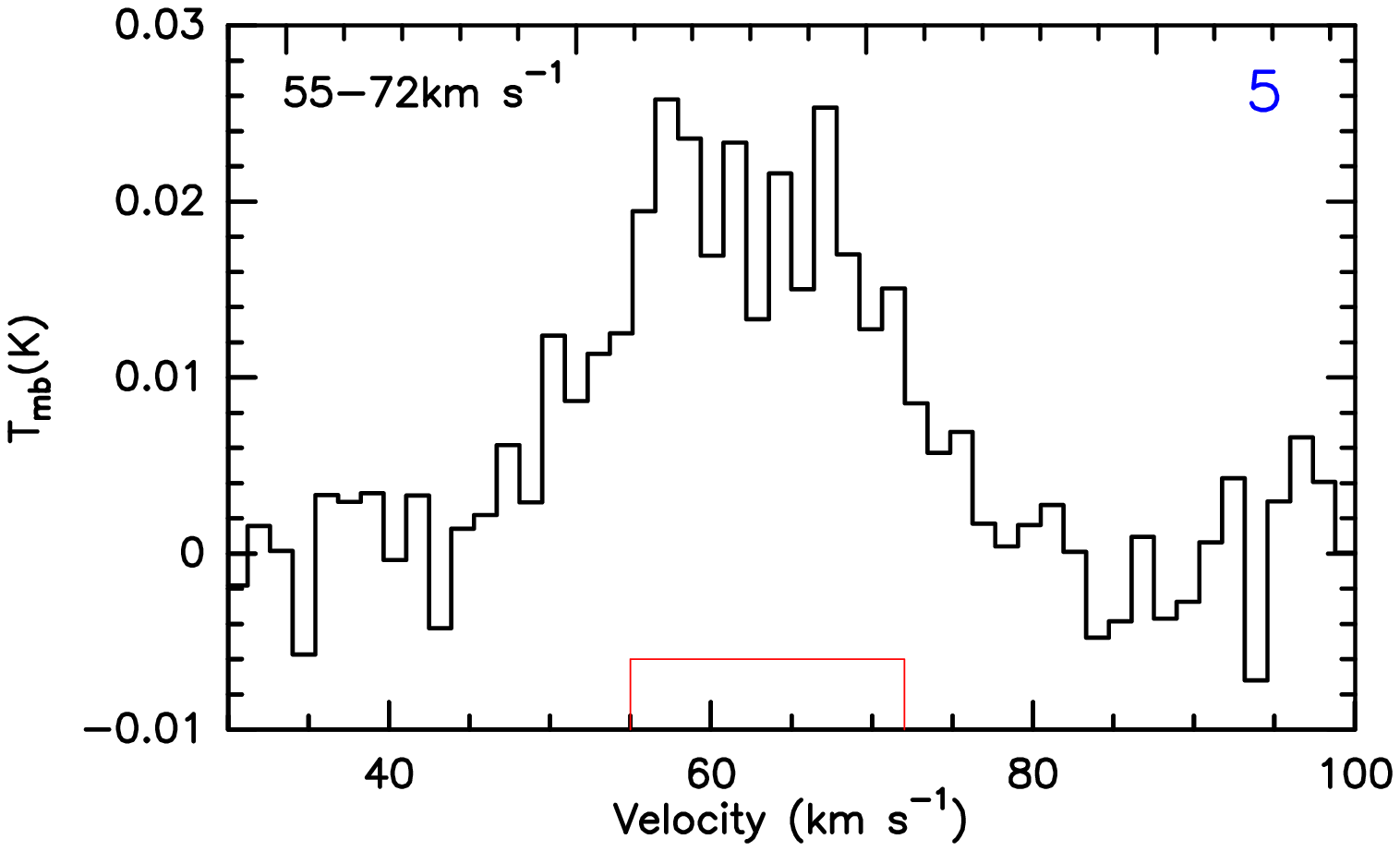}
\caption{The averaged spectra of HNC$^{18}$O. The red windows represent the velocity range used to be integrated.The numbers in the upper right corner represent the position of the region, which is shown in Figure 6.}
\label{fig 5}
\end{figure*}

\begin{figure*}
\centering
\includegraphics[width=0.47\textwidth]{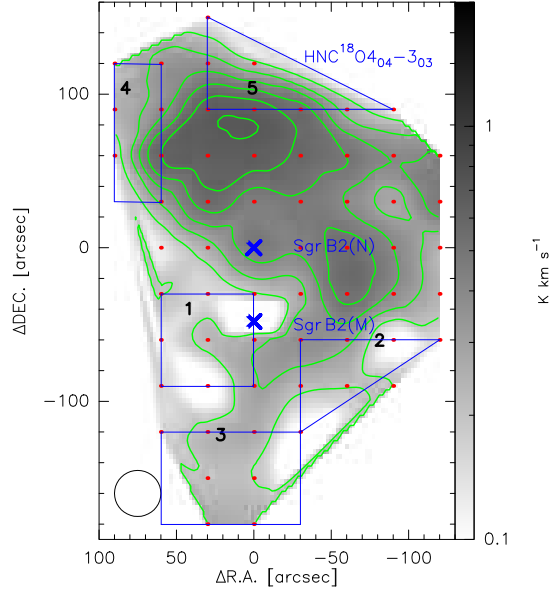}
\caption{The five regions used to calculate the averaged spectra of HNC$^{18}$O.}
\label{fig 6}
\end{figure*}

Since HN$^{13}$CO lines are strongly blended with the lines of HNCO and OC$^{34}$S, HNC$^{18}$O $4_{04}-3_{03}$ is used to calculate the optical depth of HNCO $4_{04}-3_{03}$.
%blend with HNCO and OC$^{34}$S  ! there are many HNC13O transitions
Before using the line ratio of HNC$^{18}$O and HNCO $4_{04}-3_{03}$ to obtain the optical depth of HNCO $4_{04}-3_{03}$ at each position,  the abundance ratio of HNC$^{18}$O and HNCO  needs to be known. 
   The line ratio of HNCO and HNC$^{18}$O  $4_{04}-3_{03}$ can represent the HNCO/HNC$^{18}$O abundance ratio  in the area where HNCO  $4_{04}-3_{03}$ is optically thin. The flux ratio will be the lower limit of the abundance ratio if HNCO  $4_{04}-3_{03}$ is not optically thin.   To obtain a reliable  spectrum of HNC$^{18}$O  $4_{04}-3_{03}$ with a high signal to noise ratio,  averaging data from different positions are needed.   The whole region was divided into five parts to avoid position (0,0) where Sgr B2(N) is located and the positions where the emission of HNCO  $4_{04}-3_{03}$ is optically thick. The averaged spectra are shown in Figure \ref{fig 5}, while the regions used to obtain the averaged spectra are shown in Figure \ref{fig 6}. 

The integrated flux  is used for calculating the line fluxes and ratios, which are presented in Table \ref{table 2}. The averaged ratio is shown in the last column. The flux is integrated with the same velocity range for HNCO $4_{04}-3_{03}$ and HNC$^{18}$O  $4_{04}-3_{03}$.  The red windows in Figure \ref{fig 5} are  the velocity ranges used for the integration of both HNCO and HNC$^{18}$O  $4_{04}-3_{03}$  lines,  in order to avoid  line blending  of HNC$^{18}$O from  $^{13}$CH$_2$CHCN. 
 Assuming that the HNCO/HNC$^{18}$O abundance ratio  does not vary within   Sgr B2, the HNCO/HNC$^{18}$O abundance ratio of $\sim$ 296 $\pm$ 54 is a reasonable number. The uncertainty is the square root of the sum of the variance of the HNCO/HNC$^{18}$O ratio and the square of the median of ratio error, including the uncertainties  of the measurements for the two lines. 
Assuming that the HNC$^{18}$O/HNCO abundance ratio can reflect the isotopic ratio of $^{16}$O/$^{18}$O,  it will be $\sim$ 296 $\pm$ 54. 
The result provides a more accurate value of  $^{16}$O/$^{18}$O ratio than that in the literature \citep{1980ARA&A..18..399W, 1994ARA&A..32..191W}. It is significantly different from the ratio  less than 288 derived with H$^{13}$CO$^+$/HC$^{18}$O$^+$, while  it   is still   within the uncertainty range of    406 $\pm$ 140 derived with $^{13}$CO/C$^{18}$O \citep{1980ARA&A..18..399W, 1994ARA&A..32..191W}.

\begin{table*}
\scriptsize
    \begin{center}
     \caption{The ratio between $^{16}O$ and $^{18}O$}
     \label{table 2}
     \begin{tabular}{lccc}
    \hline
    \hline
      &  $I_{HNC^{18}O}$ & $I_{HNCO}$ &  R     \\
     &  $\times$ 10$^{-2}$  K km s$^{-1}$  &  K km s$^{-1}$ &      \\
\hline
1    &  6.0(0.77)  &  14.7(0.039)  &   245(31)      \\
2    &  8.4(1.1)   &  26.1(0.14)  &   310(41)      \\
3    &  20.9(2.3)   &  47.4(0.13)   &  227(25)    \\
4    &  14.6(2.5)   &  48.2(0.064)   &  330(56)    \\
5    &  27.6(3.3)   &  73.9(0.43)   &  268(32)     \\
average  &               &               &    296(54)    \\
\hline
      \end{tabular}  \\
  \end{center}
\end{table*}

%next page
%\clearpage

\subsection{Optical depth}

With the HNCO/HNC$^{18}$O abundance ratio, the optical depth of HNCO $4_{04}-3_{03}$ toward different positions can be derived with the intensity ratio of  HNCO $4_{04}-3_{03}$ and HNC$^{18}$O $4_{04}-3_{03}$. 
%As there is contamination of $^{13}$CH$_2$CHCN, the integrated intensity of HNC$^{18}$O should be calculated after subtracting the  $^{13}$CH$_2$CHCN contribution.
As is mentioned above, the integrated intensity of HNC$^{18}$O should be calculated after subtracting the contamination of $^{13}$CH$_2$CHCN lines.
%The HNC$^{18}$O  $4_{04}-3_{03}$ at (0,0) position is blended with  one $^{13}$CH$_2$CHCN line. Therefore,  the emission of $^{13}$CH$_2$CHCN should be taken out to obtain the accurate intensity of HNC$^{18}$O.

When HNCO $4_{04}-3_{03}$ is optically thick, the optical depth can be given by the formulation:
\begin{equation}
    \frac{I_{\rm{HNCO}}}{I_{\rm{HNC^{18}O}}}=\frac{1-exp(-\tau)}{(1-exp(-\frac{\tau}{\rm{^{16}O/^{18}O}})}
\end{equation}
and the $I_{\rm{HNCO}}$ should multiple $\frac{\tau}{1-exp(-\tau)}$ after taking the optical depth into consideration.

For positions with nondetection of HNC$^{18}$O emission,  optically thin  HNCO $4_{04}-3_{03}$ is a reasonable assumption. Therefore, the intensity of HNCO will not be corrected.

\subsection{Column densities}

 The column densities of HNCO and HOCN are calculated assuming local thermodynamic equilibrium (LTE). With the integrated intensity of HNCO, the column density can be given by  \citep{2007A&A...467..207P}:

\begin{equation}
%\begin{aligned}
    N=\frac{N_J}{g_J}\times Q(T_{ex})\times exp(E_J/kT_{ex}) 
    =\frac{3h}{8\pi^3}\times \frac{1}{S\mu^2} \times \frac{I_{HNCO}\times Q(T_{ex}) }{J_{\nu}(T_{ex})-J_{\nu}(T_{bg})} \times \frac {J(T_{ex})}{\eta _{\nu}}
%\end{aligned}
\end{equation}

where $N_J$ is the column density of the upper level, $I_{HNCO}$ is the integrated intensity of HNCO after being corrected by the optical depth, $h$ is the Plank constant,  $S$ is the line strength, $\mu^2$ is the dipole moment, $\eta _{\nu}$ is the beam filling factor, $T_{ex}$ is the excitation temperature, which is equal to the rotation temperature because of the LTE hypothesis, Q(T$_{ex}$) is the partition function under the excitation temperature, J(T$_{ex}$) is defined as  $J(T_{ex})=\frac{exp(E_u/kT_{ex})}{exp(h\nu/kT_{ex})-1}$, and
J$_{\nu}$(T) is defined as  $J_{\nu}(T)=\frac{h\nu/k}{exp(h\nu/kT_{ex})-1}$. Since HNCO and HOCN emissions are extended, we assume $\eta _{\nu}$ to be 1.

To estimate the dependence of the abundant ratio on the excitation temperature, we calculate the abundance ratio of HNCO to HOCN at positions (0, 60) in different temperatures as an example. As is shown in table \ref{table 3}, the abundance ratio of HOCN to HNCO varies from 0.75\% to 0.65\% as the temperature varies from 9.375 K to 150 K, which does not change significantly.
%Therefore, %we can assume the excitation temperature to be 14 K \citep{2013A&A...559A..47B}.
Since the main goal is to obtain the relative abundance ratio of HOCN to HNCO,  14 K is a reasonable assumption for the excitation temperature. %\textbf{\citep{2013A&A...559A..47B}}.

\begin{table*}
\scriptsize
    \begin{center}
     \caption{The abundance ratio of HNCO to HOCN under different excitation temperature}
     \label{table 3}
     \begin{tabular}{lccc}
    \hline
    \hline
   Tex    &   Q(HOCN)   &  Q(HNCO)     &    R(N$_{HOCN}$/N$_{HNCO}$)   \\    
   K      &            &              &      \%    \\
\hline
9.375    &  20.1244    &  18.4492   &  0.65  \\
14       &   35.4702   &  30.6393     &   0.70   \\
18.75    &   50.8159    &  42.8291    & 0.72  \\
37.5     &    142.1413  &  117.3039   & 0.74  \\
75      &  401.3740     &   331.9879  &  0.75      \\
150      &  1135.3539   & 943.7057    &  0.75   \\

\hline
      \end{tabular}  \\
  \end{center}
Notes. Col(4): the abundance ratio of HOCN to HNCO in position (0,60).
\end{table*}

The integrated intensities are calculated using the same velocity ranges.
 The derived column densities were listed in table \ref{table 4} and table \ref{table 5}, for those positions where the emission of HOCN is higher than 5 $\sigma$ levels and lower than 5 $\sigma$ levels, respectively. 
The column density ratios between HNCO and HOCN were listed in the last column. For the positions where HOCN emission is lower than 3 $\sigma$ levels, the upper limits of the intensity and the lower limits of the ratios are given. From the results shown in table \ref{table 4} and table \ref{table 5}, the ratio of HOCN to HNCO ranges from 0.4\% to 0.7\% for most of the positions. Some positions located at the south of Sgr B2(S) have the abundance ratio of $\sim$ 0.9\%, indicating that the abundance of HOCN is enhanced. From the fourth column of table \ref{table 4} and table \ref{table 5}, we note that the optical depth of HNCO $4_{04}-3_{03}$ Sgr B2, derived from the ratio of HNCO and HNC$^{18}$O, is much smaller than 1, 
which means the emission of HNCO is almost optically thin in most of the regions of Sgr B2, while there are  just a few positions with optically thick HNCO emission. The self-absorption of HNCO lines can be omitted when dealing with the spectra of the whole area.

\begin{table}%[htb]

\begin{minipage}{100mm}
\caption{The column density of HNCO and HOCN}
\label{table 4} 	
\begin{tabular}{ccccccccc}
     \hline
     \hline
 ra & dec   &  $I_{HNCO}$ & $\tau$& $I_{HNCO}^*$ & $N_{HNCO}$ &  $I_{HOCN}$  &  $N_{HOCN}$  &  R  \\
 \arcsec  &  \arcsec   & K km s$^{-1}$ &    &  K km s$^{-1}$ & $\times$ 10$^{15}$ cm$^{-2}$ &  K km s$^{-1}$ & $\times$ 10$^{13}$ cm$^{-2}$ &  \%  \\
\hline
0    & 0    & 163.59(0.81) & 0    & 163.59 & 2.23 & 3.92(0.10) & 1.14 & 0.50 \\
60   & 60   & 211.89(0.44) & 0.26 & 240.63 & 3.27 & 7.15(0.06) & 2.08 & 0.63 \\
-60  & 60   & 161.98(0.45) & 0.15 & 174.43 & 2.37 & 4.84(0.05) & 1.40 & 0.58 \\
60   & -60  & 75.68(0.32)  & 0    & 75.68  & 1.03 & 2.04(0.06) & 0.59 & 0.57 \\
-60  & -60  & 140.66(0.98) & 0    & 140.66 & 1.91 & 4.38(0.05) & 1.27 & 0.66 \\
30   & 0    & 85.05(0.49)  & 0.14 & 91.14  & 1.24 & 2.30(0.04) & 0.67 & 0.53 \\
-30  & 0    & 196.58(1.42) & 0    & 196.58 & 2.67 & 6.80(0.04) & 1.97 & 0.73 \\
0    & 30   & 216.54(0.55) & 0    & 216.54 & 2.95 & 7.06(0.04) & 2.05 & 0.69 \\
0    & -30  & 125.53(0.77) & 0    & 125.53 & 1.71 & 3.07(0.04) & 0.89 & 0.51 \\
-30  & 30   & 213.91(0.84) & 0    & 213.91 & 2.91 & 7.49(0.04) & 2.17 & 0.74 \\
-30  & -30  & 223.97(1.83) & 0    & 223.97 & 3.05 & 7.40(0.05) & 2.15 & 0.70 \\
30   & 30   & 217.95(0.42) & 0    & 217.95 & 2.97 & 7.09(0.04) & 2.06 & 0.68 \\
30   & -30  & 73.33(0.47)  & 0    & 73.33  & 1.00 & 1.99(0.05) & 0.58 & 0.57 \\
0    & 60   & 257.17(0.59) & 0    & 257.17 & 3.50 & 8.71(0.05) & 2.53 & 0.71 \\
60   & 0    & 66.58(0.44)  & 0    & 66.58  & 0.91 & 1.62(0.06) & 0.47 & 0.51 \\
-60  & 0    & 177.18(1.72) & 0    & 177.18 & 2.41 & 5.72(0.05) & 1.66 & 0.68 \\
0    & -60  & 114.15(0.56) & 0    & 114.15 & 1.55 & 3.83(0.06) & 1.11 & 0.71 \\
30   & 60   & 265.35(0.23) & 0.03 & 269.35 & 3.66 & 9.63(0.04) & 2.80 & 0.75 \\
-30  & 60   & 175.12(0.29) & 0.53 & 225.61 & 3.07 & 5.70(0.05) & 1.66 & 0.53 \\
60   & 30   & 132.75(0.51) & 0    & 132.75 & 1.81 & 4.00(0.03) & 1.16 & 0.63 \\
-60  & 30   & 191.85(1.10) & 0.35 & 227.38 & 3.09 & 6.53(0.04) & 1.90 & 0.60 \\
-60  & -30  & 189.09(2.04) & 0    & 189.09 & 2.57 & 5.83(0.06) & 1.69 & 0.65 \\
30   & -60  & 68.57(0.28)  & 0    & 68.57  & 0.93 & 1.95(0.04) & 0.57 & 0.60 \\
-30  & -60  & 188.86(1.26) & 0    & 188.86 & 2.57 & 6.93(0.07) & 2.01 & 0.77 \\
-90  & 0    & 126.98(1.44) & 0    & 126.98 & 1.73 & 2.85(0.05) & 0.83 & 0.47 \\
0    & 90   & 216.62(0.31) & 0.14 & 232.14 & 3.16 & 6.13(0.04) & 1.78 & 0.56 \\
0    & -90  & 120.38(0.32) & 0    & 120.38 & 1.64 & 5.40(0.03) & 1.57 & 0.94 \\
60   & 90   & 182.59(0.47) & 0    & 182.59 & 2.48 & 3.93(0.07) & 1.14 & 0.44 \\
30   & 90   & 217.89(0.24) & 0.17 & 236.94 & 3.22 & 5.51(0.06) & 1.60 & 0.49 \\
-30  & 90   & 180.20(0.19) & 0.09 & 188.43 & 2.56 & 5.12(0.06) & 1.49 & 0.57 \\
-60  & 90   & 130.04(0.23) & 0    & 130.04 & 1.77 & 2.86(0.05) & 0.83 & 0.46 \\
-90  & 90   & 103.41(0.12) & 0    & 103.41 & 1.41 & 1.94(0.04) & 0.56 & 0.39 \\
60   & -90  & 81.82(0.13)  & 0    & 81.82  & 1.11 & 2.11(0.06) & 0.61 & 0.54 \\
30   & -90  & 72.36(0.32)  & 0    & 72.36  & 0.98 & 2.77(0.04) & 0.80 & 0.81 \\
-30  & -90  & 145.57(0.74) & 0    & 145.57 & 1.98 & 5.56(0.05) & 1.62 & 0.80 \\
-60  & -90  & 96.98(0.68)  & 0    & 96.98  & 1.32 & 2.70(0.05) & 0.78 & 0.59 \\
-90  & -90  & 76.38(0.90)  & 0    & 76.38  & 1.04 & 1.98(0.05) & 0.57 & 0.54 \\
-90  & 30   & 159.60(1.02) & 0    & 159.60 & 2.17 & 3.85(0.05) & 1.12 & 0.51 \\
\noalign{\smallskip}\hline
\end{tabular}
\end{minipage}
\end{table}

\addtocounter{table}{-1}
\begin{table}%[htb]

\begin{minipage}{100mm}
\caption{$-$ Continued.}
\label{table 4} 	
\begin{tabular}{ccccccccc}
\hline
\hline
ra & dec   &  $I_{HNCO}$ & $\tau$& $I_{HNCO}^*$ & $N_{HNCO}$ &  $I_{HOCN}$  &  $N_{HOCN}$  &  R  \\
 \arcsec  &  \arcsec   & K km s$^{-1}$ &    &  K km s$^{-1}$ & $\times$ 10$^{15}$ cm$^{-2}$ &  K km s$^{-1}$ & $\times$ 10$^{13}$ cm$^{-2}$ &  \%  \\
\hline

-90  & 60   & 135.04(0.45) & 0.05 & 138.44 & 1.88 & 3.10(0.05) & 0.90 & 0.47 \\
-120 & 0    & 131.15(0.59) & 0    & 131.15 & 1.78 & 2.35(0.05) & 0.68 & 0.38 \\
-120 & 30   & 147.17(0.38) & 0    & 147.17 & 2.00 & 2.69(0.05) & 0.78 & 0.38 \\
-120 & 60   & 119.15(0.20) & 0    & 119.15 & 1.62 & 2.20(0.05) & 0.64 & 0.39 \\
-90  & -30  & 118.62(1.73) & 0.22 & 132.15 & 1.80 & 2.82(0.05) & 0.82 & 0.45 \\
-90  & -60  & 96.55(1.22)  & 0    & 96.55  & 1.31 & 2.40(0.05) & 0.70 & 0.52 \\
0    & 120  & 126.35(0.22) & 0    & 126.35 & 1.72 & 2.21(0.04) & 0.64 & 0.37 \\
30   & 120  & 127.80(0.18) & 0    & 127.80 & 1.74 & 2.50(0.07) & 0.72 & 0.41 \\
90   & 90   & 127.30(0.52) & 0    & 127.30 & 1.73 & 2.66(0.08) & 0.77 & 0.44 \\
90   & 60   & 134.89(0.64) & 0    & 134.89 & 1.84 & 2.67(0.06) & 0.77 & 0.42 \\
90   & 120  & 46.97(0.11)  & 0    & 46.97  & 0.64 & 0.90(0.03) & 0.26 & 0.40 \\
-30  & -120 & 87.69(0.94)  & 0    & 87.69  & 1.19 & 3.16(0.06) & 0.92 & 0.76 \\
0    & -120 & 126.08(0.42) & 0    & 126.08 & 1.72 & 5.13(0.06) & 1.49 & 0.86 \\
30   & -120 & 85.06(0.15)  & 0    & 85.06  & 1.16 & 3.70(0.05) & 1.08 & 0.92 \\
60   & -120 & 78.10(0.21)  & 0    & 78.10  & 1.06 & 1.90(0.05) & 0.55 & 0.51 \\
0    & -150 & 100.92(0.35) & 0    & 100.92 & 1.37 & 3.68(0.04) & 1.07 & 0.77 \\
30   & -150 & 88.28(0.25)  & 0    & 88.28  & 1.20 & 3.59(0.06) & 1.04 & 0.86 \\
0    & -180 & 131.58(0.77) & 0    & 131.58 & 1.79 & 4.34(0.07) & 1.26 & 0.69 \\
30   & -180 & 78.01(0.27)  & 0    & 78.01  & 1.06 & 2.98(0.04) & 0.87 & 0.80 \\
				
\noalign{\smallskip}\hline
\end{tabular}
\end{minipage}

Note. Col(1) and Col(2): the equatorial offsets of emission with respect to Sgr B2(N); Col(3): the integrated intensity of HNCO, 1 $\sigma$ level error is given; Col(4): the optical depth of HNCO; Col(5): the integrated intensity of HNCO after considering the optical depth; Col(6): the column density of HNCO; Col(7): the integrated intensity of HOCN, 1 $\sigma$ level error is given; Col(8): the column density of HOCN; Col(9): the abundance ratio of HOCN to HNCO.  
\end{table}

\begin{table*}
\scriptsize
    \begin{center}
     \begin{minipage}{100mm}
     \caption{The column density of HNCO and HOCN}
     \label{table 5}  
     \begin{tabular}{ccccccccc}
     \hline
     \hline
 ra & dec   &  $I_{HNCO}$ & $\tau$& $I_{HNCO}^*$ & $N_{HNCO}$ &  $I_{HOCN}$  &  $N_{HNCO}$  &  R  \\
  \arcsec  &  \arcsec   & K km s$^{-1}$ &    &  K km s$^{-1}$ & $\times$ 10$^{15}$ cm$^{-2}$ &  K km s$^{-1}$ & $\times$ 10$^{13}$ cm$^{-2}$ &   \%  \\
\hline

-120 & -30  & 82.38(1.10)  & 0    & 82.38  & 1.12 & 1.38(0.05) & 0.40 & 0.35 \\
60   & -30  & 54.11(0.40)  & 0    & 54.11  & 0.74 & 1.50(0.05) & 0.43 & 0.58 \\
-120 & -60  & 63.88(0.72)  & 0    & 63.88  & 0.87 & 1.14(0.05) & 0.33 & 0.37 \\
60   & 120  & 110.55(0.25) & 0    & 110.55 & 1.50 & 1.93(0.08) & 0.56 & 0.37 \\
30   & 150  & 74.51(0.25)  & 0    & 74.51  & 1.01 & 0.76(0.07) & 0.22 & 0.22 \\
0    & -48  & 110.56(0.64) & 0    & 110.56 & 1.50 & 2.76(0.08) & 0.80 & 0.53 \\
\hline
      \end{tabular}     
    \end{minipage}
   \end{center}

 Note. The positions shown in table \ref{table 5} are the area where the emission of HOCN is lower than 5 $\sigma$ levels. Col(1) and Col(2): the equatorial offsets of emission with respect to Sgr B2(N); Col(3): the integrated intensity of HNCO, 1 $\sigma$ level error is given; Col(4): the optical depth of HNCO; Col(5): the integrated intensity of HNCO after considering the optical depth; Col(6): the column density of HNCO; Col(7): the integrated intensity of HOCN, 1 $\sigma$ level error is given; Col(8): the column density of HOCN; Col(9): the abundance ratio of HOCN to HNCO.
\end{table*}

\section{Discussion}
\label{sect:discussion}
%We use the IRAM 30m to map HNCO and HOCN towards Sgr B2.  In the current observation,
HOCN emission was found to  exist in most positions of Sgr B2, confirming that HOCN is also widespread in Sgr B2,  just like its isomer specie HNCO. The similarity between the map of HNCO and HOCN emission indicates that they may have common origins. The strong emission discovered in the north of the cloud is likely to be related to the large-scale shock that exists in Sgr B2. The abundance ratio of HOCN to HNCO ranges from 0.4\% to 0.7\%  in most of the positions. The histogram of the abundance ratio of HOCN to HNCO is shown in Figure \ref{fig 7}.

\subsection{Isotopic ratio}

The isotopic ratio of oxygen gained from HNCO and HNC$^{18}$O is 296 $\pm$ 54. The assumption used during the calculation  is that there is no chemical priority, namely that the ratio of HNCO/HNC$^{18}$O can represent the ratio of $^{16}$O/$^{18}$O. The validity of the assumption needs to be confirmed, as strong chemical priority will change our result seriously. 
In addition, the intensity of HNC$^{18}$O also plays an important role in the result. 
As shown in Figure \ref{fig 5}, the intensity of HNC$^{18}$O after being averaged is just slightly a little bit stronger than 3$\sigma$, while the baseline is hard to be corrected. So the intensities of HNC$^{18}$O can not be well determined. %The isotopic  ratio of  296 $\pm$ 43 need be confirmed by other methods. 

\subsection{Abundance ratio of HNCO and HOCN}
%There are 10 positions \textbf{with the abundance ratio of} HOCN to HNCO smaller than 0.4\% and 7 positions larger than 0.8\%

\begin{figure*}
\centering
\includegraphics[width=0.47\textwidth]{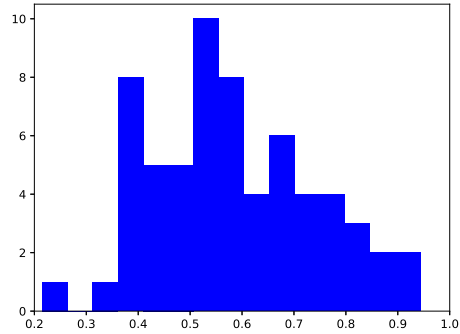}
\caption{The histogram of the abundance ratios}
\label{fig 7}
\end{figure*}

\begin{figure*}
\centering
\includegraphics[width=0.47\textwidth]{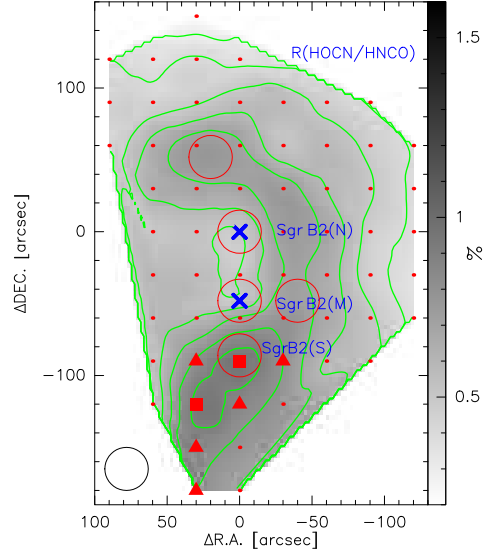}
\caption{The spatial distribution of HOCN to HNCO ratio. The contour levels range from 30\% $\sim$ 100\% with the step of 10\% and the peak value is 0.94\%, while the greyscale is this ratio in percentage.
The red triangles and red squares mark the positions where the abundance ratio of  HOCN to HNCO is larger than 0.8\%. The red squares show the positions with the largest abundance ratio. The red circles represent the positions where \cite{2010A&A...516A.109B} have observed. The red circle under Sgr B2(M) is Sgr B2 (S).}
\label{fig 8}
\end{figure*}

\begin{figure*}
\centering
\includegraphics[width=0.4\textwidth]{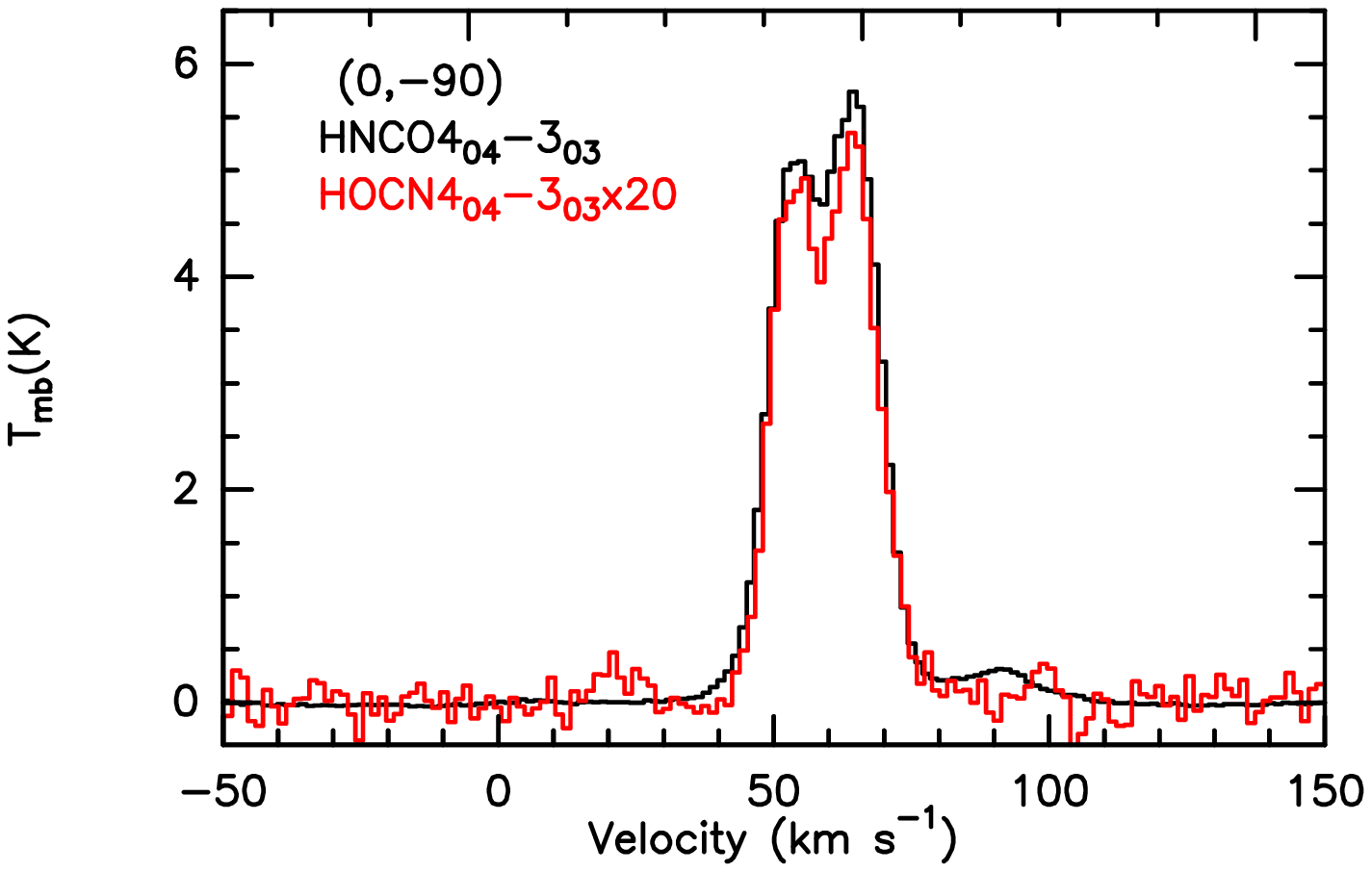}
\includegraphics[width=0.4\textwidth]{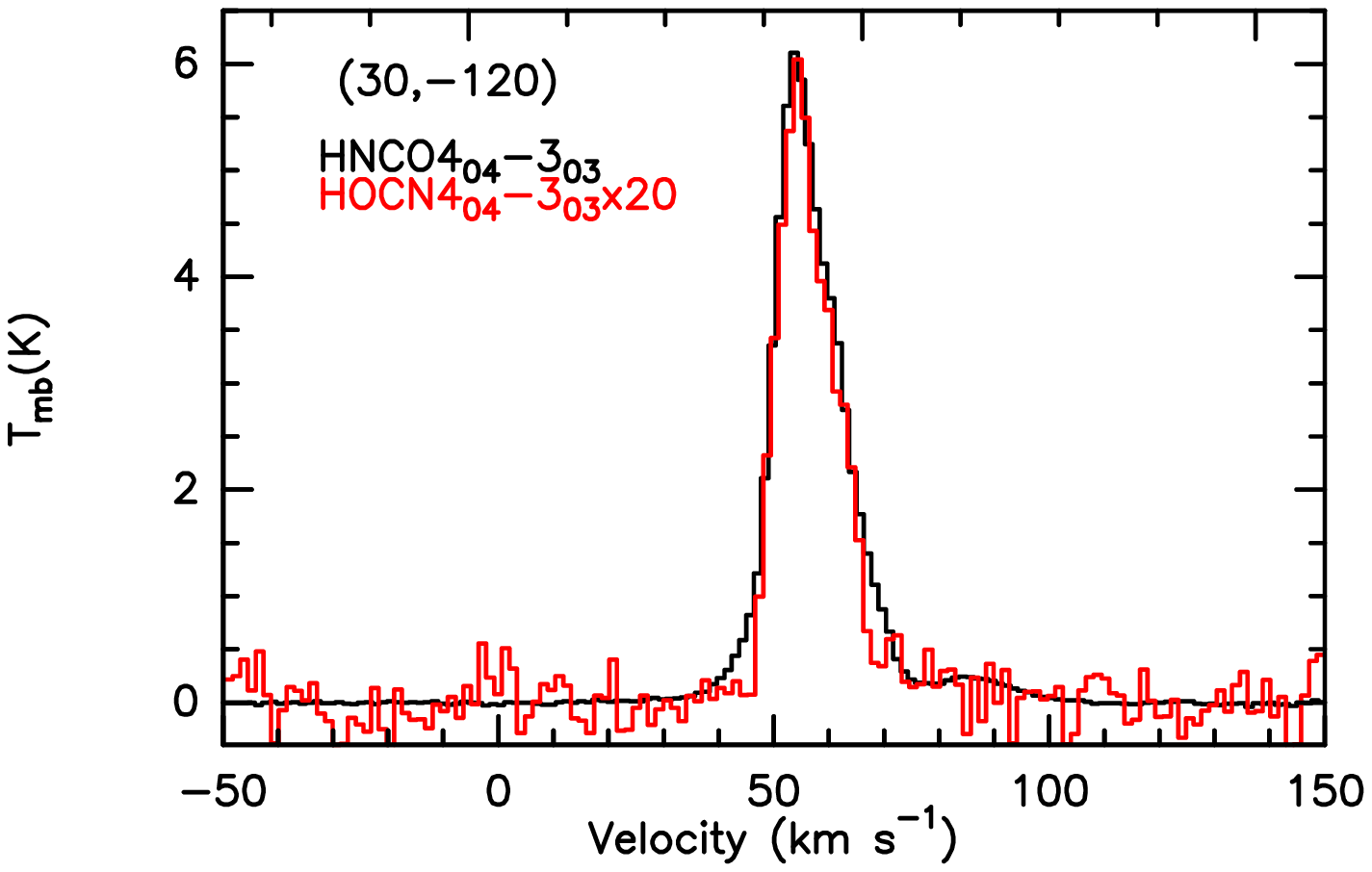}
\caption{The spectra of HNCO $4_{04}-3_{03}$ and HOCN $4_{04}-3_{03}$ in (0, -90) and (30, -150), marked as red squares  in Figure \ref{fig 8}.}
\label{fig 9}
\end{figure*}

The abundance ratio of HOCN to HNCO is smaller than 0.4\% toward 10 positions, while this value is larger than 0.8\% toward 7 positions (see Figure \ref{fig 7}).
All the 10 positions with extremely low HOCN to HNCO abundance ratio are located in the boundary of  this map, with weak HOCN lines which cause large uncertainties of the abundance  ratio. On the other hand,  the seven positions with the abundance ratios higher than 0.8\%   located around Sgr B2(S) (marked as red triangles or squares in Figure \ref{fig 8}), are with  accurate values. The non-detection of HNC$^{18}$O $4_{04}-3_{03}$ toward these positions provides small optical depth of  HNCO $4_{04}-3_{03}$,  which means the underestimation of HNCO is not important. The spectra of HNCO $4_{04}-3_{03}$ and HOCN $4_{04}-3_{03}$ toward positions (0, -90) and (30, -150) are shown in Figure \ref{fig 9} as an example. The high signal to noise ratio ensures that the abundance of HOCN is actually enhanced in these positions,  which needs to be explained by  new chemical models.
The ratios in the rest regions  range from 0.4\% to 0.7\%, without significant variation.
% HNCO is not weak

The strong continuum in Sgr B2(N) and (M) may also have a contribution to the line emission of HNCO. This is ignored to simplify our calculation, so the abundance ratio in Sgr B2(N) and (M) need to be updated in the future.

%From the column densities calculated, we note that the abundance ratio of HNCO and HOCN are lower in the boundary of Sgr B2 than the other points. This result could be attributed to the temperature distribution in Sgr B2. Since the energy used during the formation of HNCO is lower than HOCN  \citep{2004JChPh.12011586S}, which allows it to form more effective than HOCN in the low temperature environment. So the larger ratio may indicate that the temperature in the area which is away from the hot core is low. In addition, the emission of HOCN is not very strong in the boundary of Sgr B2. In this case, the signal to noise ratio is low, resulting in that the integrated intensity of HOCN can not be well calculated in those position. This is another reason that may account for the smaller abundance ratio. One the other hand, the absorption energy for HOCN is about two times larger than HNCO  \citep{2015A&A...578A..62L}, which means more HOCN will be released from the grain in the region where temperature is higer than other regions. This factor will also play a role on the ratio of HOCN and HNCO. 

\subsection{Chemical models}

%After learning the abundance distribution of HNCO and HOCN, we can have a discussion about the chemical models for HOCN. 

The obtained abundance ratios of HOCN/HNCO  are about 0.4\% to 0.7\% in most regions of Sgr B2, implying that the formation mechanism of these two molecules does not vary  in different parts of  Sgr B2 complex. This result agrees well  with the calculated result of the gas-grain model  \citep{2010ApJ...725.2101Q}. In this model,  the formation and destruction of both molecules involves gas-phase reactions and grain-surface reactions. HNCO and HOCN initially form on the grain surface, and then they are released from the grain by shocks or some other processes. The fact that the abundance of HOCN is enhanced around Sgr B2(S) needs new chemical models to explain it.

Further observation needs to be conducted to get a more detailed understanding of HNCO and its other isomer in other different sources. It will give us more insights into how the interstellar environment affect the relative abundance of isomer families.

\section{Summary}

With point-by-point mapping observations of HOCN and HNCO lines around Sgr B2 with IRAM 30m telescope,  the spatial distribution of HNCO 4$_{04}$-3$_{03}$, HOCN 4$_{04}$-3$_{03}$, HNC$^{18}$O and HNCO 4$_{14}$-3$_{13}$ were obtained. Our main results include:

1. From the spatial distribution of HOCN which is similar to the one of HNCO, we can refer that HOCN is extended in Sgr B2, and perhaps, has a close relationship with HNCO.

2. We note that HOCN molecule is enhanced around Sgr B2(S), with the HOCN to HNCO abundance ratio of  $\sim$ 0.9\%, while this ratio changes little in the rest positions, ranging from  0.4\% to 0.7\%. Given the relatively constant abundance ratio of 0.4\% to 0.7\%, which agrees with the gas-grain model well \citep{2010ApJ...725.2101Q}, the formation mechanism may be involved both gas-phase reaction and grain-surface reaction, not varying a lot in most parts of Sgr B2.

3. The isotopic ratio of $^{16}$O/$^{18}$O derived from the ratio of HNCO/HNC$^{18}$O is 296 $\pm$ 54 in Sgr B2. The optical depths of HNCO  in most regions of Sgr B2 derived from HNCO/HNC$^{18}$O line ratio are  smaller than 1, indicating that HNCO 4$_{04}$-3$_{03}$ is almost optically thin there. 

Whether the condition in Sgr B2 would also apply to other sources may also be a question that  needs to be addressed, with further large sample surveys. % If the result still hold, the relationship between HNCO and HOCN could be clarify better, and it may also give some insight into the research of other isomeric families. 

\normalem
\begin{acknowledgements}
The authors thank the staff at IRAM for their excellent support of these observations. This work made use of the CDMS Database. This work has been supported by the Natural Science Foundation of China (11773054 and U1731237). This work is also supported by the international partnership program of Chinese Academy of Sciences through grant No.114231KYSB20200009. The single dish data are available in the IRAM archive at https://www.iram-institute.org/EN/content-page-386-7-386-0-0-0.html.

\end{acknowledgements}
  
\bibliographystyle{raa}
\bibliography{bibtex}

\end{document}